\documentclass[sigconf]{acmart}

\usepackage{amsthm}
\usepackage{amsmath}
\usepackage{times}
\usepackage{kotex}
\usepackage{soul}
\usepackage{xcolor}
\usepackage{url}
\usepackage{enumitem}
\usepackage{comment}
\usepackage{threeparttable}
\usepackage[switch]{lineno}
\usepackage{lipsum}
\usepackage[linesnumbered,algoruled,boxed,lined,noend]{algorithm2e}
\usepackage{graphicx,subcaption}
\usepackage{fancyhdr} 
\usepackage[utf8]{inputenc}
\usepackage{multirow}
\usepackage[normalem]{ulem}
\usepackage{multicol}
\usepackage{makecell}
\usepackage{cancel}
\usepackage{cellspace}
\usepackage{fontawesome}
\usepackage{microtype}
\usepackage{float}
\usepackage{placeins} 
\usepackage{stfloats}
\DeclareMathOperator*{\argmin}{arg\,min}

\newcommand{\spara}[1]{\smallskip\noindent{\bf #1}}

\newtheorem{definition}{Definition}
\newtheorem{property}{Property}

\newtheorem{example}{Example}

\AtBeginDocument{%
  }

\setcopyright{none}
\copyrightyear{2018}
\acmYear{2018}
\acmDOI{XXXXXXX.XXXXXXX}
\acmConference[Conference acronym 'XX]{Make sure to enter the correct
  conference title from your rights confirmation email}{June 03--05,
  2018}{Woodstock, NY}

\acmISBN{978-1-4503-XXXX-X/2018/06}

\begin{document}
\fancyfoot{}

\makeatletter
\def\fps@figure{tbp}
\def\fps@figure*{tbp} 
\makeatother

\title{Experimental Analysis and Evaluation of Cohesive Subgraph Discovery}

\author{Dahee Kim}
\affiliation{%
  \institution{UNIST}
  \city{Ulsan}
  \country{Republic of Korea}
}
\email{dahee@unist.ac.kr}

\author{Song Kim}
\affiliation{%
  \institution{UNIST}
  \city{Ulsan}
  \country{Republic of Korea}
}
\email{song.kim@unist.ac.kr}

\author{Jeongseon Kim}
\affiliation{%
  \institution{Chungnam National University}
  \city{Daejeon}
  \country{Republic of Korea}
}
\email{jeongseon@cnu.ac.kr}

\author{Junghoon Kim}
\affiliation{%
  \institution{UNIST}
  \city{Ulsan}
  \country{Republic of Korea}
}
\email{junghoon.kim@unist.ac.kr}

\author{Kaiyu Feng}
\affiliation{%
  \institution{Beijing Institute of Technology}
  \city{Beijing}
  \country{China}
}
\email{fengky@bit.edu.cn}

\author{Sungsu Lim}
\affiliation{%
  \institution{Chungnam National University}
  \city{Daejeon}
  \country{Republic of Korea}
}
\email{sungsu@cnu.ac.kr}

\author{Jungeun Kim}
\affiliation{%
  \institution{Kongju National University}
  \city{Cheonan}
  \country{Republic of Korea}
}
\email{jekim@kongju.ac.kr}

\renewcommand{\shortauthors}{Kim et al.}

\begin{abstract}
Retrieving cohesive subgraphs in networks is a fundamental problem in social network analysis and graph data management. These subgraphs can be used for marketing strategies or recommendation systems. 
Despite the introduction of numerous models over the years, a systematic comparison of their performance, especially across varied network configurations, remains unexplored. 
In this study, we evaluated various cohesive subgraph models using task-based evaluations and conducted extensive experimental studies on both synthetic and real-world networks. Thus, we unveil the characteristics of cohesive subgraph models, highlighting their efficiency and applicability. 
Our findings not only provide a detailed evaluation of current models but also lay the groundwork for future research by shedding light on the balance between the interpretability and cohesion of the subgraphs. This research guides the selection of suitable models for specific analytical needs and applications, providing valuable insights.
\end{abstract}

\maketitle

\def\thefootnote{*}
\footnotetext{Corresponding author}
\def\thefootnote{\arabic{footnote}}  

\def\thefootnote{$+$ }
\footnotetext{This is the accepted version of the paper published in Information Sciences (Elsevier).
The final published version is available at: \url{https://doi.org/10.1016/j.ins.2024.120664}}

\section{Introduction}
The proliferation of internet services and mobile devices has made it possible for people to access social networking services anytime and anywhere. As a result, the scale of social networks has increased rapidly. For instance, Facebook, one of the most popular social networking services in the world, had over 2.93 billion monthly active users as of the first quarter of 2022. This availability of large-scale social networks has encouraged research into social network analysis, which aims to understand human behaviour and various social phenomena by analysing social networks. Social network analysis has garnered significant attention from various fields~\cite{barabasi2013network}.

Among the various topics in network science~\cite{fu2024community, barabasi2013network}, discovering cohesive subgraphs is a fundamental problem with wide-ranging applications~\cite{wang2024efficient}. The goal of cohesive subgraph discovery is to find a subgraph of a given graph that is highly connected or cohesive. It has many applications including graph visualisation~\cite{batagelj1999partitioning}, community search~\cite{fang2020survey,sozio2010community}, key groups~\cite{seo2021finding}, keyword extraction~\cite{rousseau2015main}, and more.

Several cohesive subgraph models have been proposed in recent years, which can be categorised into the following types based on the graph structure used to define the subgraph: (1) core-based models~\cite{seidman1983network,zhang2018discovering,zhang2020exploring,bonchi2019distance,govindan2017k}, (2) triangle-based models~\cite{cohen2008trusses, wu2019towards}, (3) clique-based models~\cite{luce1949method, wasserman1994social, tang2010community, alba1973graph, mokken1979cliques}, (4) connected component-based models~\cite{wang2015simple}, and (5) other types of models~\cite{victor2021alphacore, li2018discovering}. Each of these models addresses the limitations of other models and aims to capture high-level cohesiveness based on its interest. However, this complexity makes it difficult to determine which cohesive model performs best in specific types of networks or scenarios. Furthermore, as cohesive subgraph discovery is a fundamental step in many applications, obtaining a general overview of the performance of existing cohesive subgraph models in various downstream tasks is challenging. Therefore, there is a clear need for an empirical study that benchmarks the performance of existing cohesive subgraph models and provides in-depth insight into the relative merit of these models and their applicable scenarios.

To the best of our knowledge, this paper is the first to provide a comprehensive evaluation of various cohesive subgraph models through experimental studies and interpretation. The evaluation includes fourteen existing cohesive subgraph models, twelve representative real-world graph datasets, LFR synthetic datasets, seven evaluation metrics, and a downstream task.

We ensure that the above evaluation will offer insights into which cohesive subgraph models perform better for different scenarios. This paper provides the first comprehensive evaluation of $14$ existing cohesive subgraph models through experimental studies and interpretation, with a focus on simple undirected and unweighted graphs. Note that there are numerous cohesive subgraph discovery approaches in various network settings, such as signed networks~\cite{kim2023effective}, hypergraphs~\cite{kgcore},  bipartite networks~\cite{abc, he2021exploring}, multi-layer graph~\cite{liu2023gcore}, and more. 

The increase in research across various networks has arisen from advancements in large-scale data analysis capabilities. These efforts are evolving to efficiently manage and extend information from traditional graph structures to more complex network settings.

In this paper, we mainly focus on simple undirected and unweighted graphs due to its fundamental and simple structure. By studying cohesive subgraph models in this setting, we aim to establish a foundation and provide insights, paving the way for future research to explore more specialised network types.

\begin{figure}[t]
\centering
\includegraphics[width=0.55\linewidth]{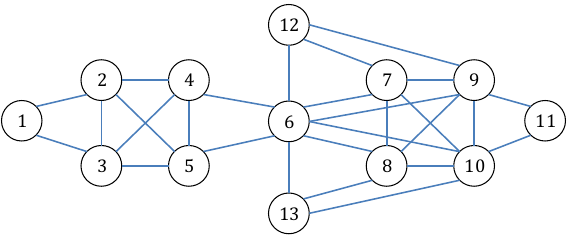}
\vspace{-0.1cm}
        \caption{Simple toy network}
        \label{fig:main_example} 
\end{figure}

\spara{Contribution.} Our contributions are summarised as follows:

\begin{itemize}[leftmargin=*]
\item \textbf{Unifying and implementing existing algorithms}: This is the first work to implement existing cohesive subgraph models and combine them into a single project. We offer the public source code online for further research.
\item \textbf{Experimental study on large-sized datasets with different scenarios}: We report extensive experimental results from various perspectives using real-world and synthetic networks with different configurations. With diverse downstream tasks based on the cohesive subgraph models, we also suggest a suitable cohesive subgraph model. 
\item \textbf{Insights into model performance across network configuration}: Our study delivers a focused analysis on cohesive subgraph models across different network configurations. It highlights key characteristics and guides the selection of models for specific research purposes, laying groundwork for future explorations in network analysis.

\end{itemize}

The rest of this paper is organised as follows. Section~\ref{sec:preliminaries} describes the fundamentals of the cohesive subgraph discovery problem. Section~\ref{sec:models} surveys cohesive subgraph models. Section~\ref{sec:evaluation} evaluates the models and reports the evaluation results. Section~\ref{sec:challenges} discusses the challenges for future research.
Finally, Section~\ref{sec:conclusion} 
concludes and summarises the key findings.

\section{Preliminaries}\label{sec:preliminaries}
In this section, we discuss the preliminaries for this paper. 
Table~\ref{tab:notations} presents the basic notations which are used in the overall paper. 
We consider an input graph to be unweighted, simple, and undirected. 

\begin{figure}[t]
\centering
\includegraphics[width=0.55\linewidth]{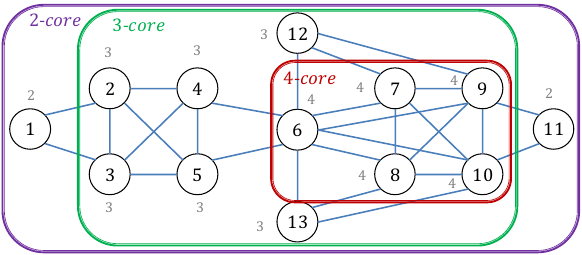}
\vspace{-0.1cm}
        \caption{Core decomposition, computation}
        \label{fig:decom_com} 
\end{figure}

The notation \textit{decomposition} is the process that obtaining information about all cohesive subgraphs in a graph. 
The \textit{computation} aims to focus on finding a specific cohesive subgraph based on user-defined parameters.
For example, $k$-core \textit{decomposition}  and \textit{computation} can be considered. In $k$-core \textit{decomposition}, we can obtain the coreness values of all nodes in a graph. For each node, a coreness value is   $k$ if it belongs to the $k$-core but not to $(k+1)$-core. Given a user parameter $k'$, we can simply find $k'$-core by identifying a set of nodes of which its coreness value is larger than or equal to $k'$. 
In $k$-core \textit{computation}, we aim to identify a cohesive subgraph that satisfies a specific parameter $k$ by iteratively removing a set of nodes which does not meet the degree constraint. 
Figure~\ref{fig:decom_com} shows the $k$-core and coreness values for each node in Figure~\ref{fig:main_example}.

\begin{table}[hbt!]
\centering
\footnotesize
\caption{Notations}
\vspace{-0.2cm}
\label{tab:notations}
\centering
\begin{tabular}{c||l}
\hline
\textbf{Notation} & \textbf{Description} \\ \hline \hline
$G=(V,E)$& a network\\ \hline
$G[C]$ & an induced subgraph by $C$ in $G$ \\ \hline
$V[C]; C$ & a set of nodes $C$  \\ \hline
$E[C]$ & a set of edges in $G[C]$ \\ \hline
$k$     & structural threshold    \\ \hline
$h$     & distance threshold for $(k,h)$-core~\cite{bonchi2019distance}\\ \hline
$s$     & strong tie threshold for $(k,s)$-core~\cite{zhang2018discovering}\\ \hline
$p$     & fraction threshold for $(k,p)$-core~\cite{zhang2020exploring}\\ \hline
$\alpha$    & threshold for Alphacore~\cite{victor2021alphacore} and $k$-core-truss~\cite{li2018discovering} \\ \hline
$\epsilon$    & similarity threshold for SCAN~\cite{xu2007scan} \\ \hline
$sup(e)$& support of an edge $e$, i.e., $\#$ of triangles containing $e$ \\ \hline
$d(u); d(u,G)$      & a degree of node $u$ in $G$      \\ \hline
$d(u, G')$& a degree of node $u$ in $G'$      \\ \hline
$diam(G')$& a diameter of a graph $G'$      \\ \hline
$dist(u,v)$    & a distance from $u$ to $v$      \\ \hline
$dist(u,v,G')$    & a distance from $u$ to $v$ in $G'$      \\ \hline
$\delta(G')$    & minimum degree of $G'$, i.e., $\argmin_{v\in V[G']} d(v, G')$      \\ \hline
$\Delta_{u,v,w}$    & a triangle formed by nodes $u$, $v$, and $w$      \\ \hline
$\Delta(u)$ & a set of triangles containing node $u$ \\ \hline
\hline
\end{tabular}
\end{table}

\begin{table}[t]
\caption{Algorithm notation}
\vspace{-0.2cm}
\footnotesize
\label{tab:alg_notation}
\centering
\begin{tabular}{c|c|c}
\hline \hline
\textbf{Algorithm}        & \textbf{Notation}   & \textbf{Category}
\\ \hline \hline
$k$-core~\cite{seidman1983network}& $C^{kc}_k$  & Core-based  \\ \hline
$(k,h)$-core~\cite{bonchi2019distance}       & $C^{khc}_{k,h}$ & Core-based \\ \hline
$(k,p)$-core\cite{zhang2020exploring}       & $C^{kpc}_{k,p}$ & Core-based \\ \hline
$k$-peak~\cite{govindan2017k} & $C^{kp}_{k}$ & Core-based \\ \hline
$k$-truss~\cite{cohen2008trusses} & $C^{kt}_{k}$ & Triangle-based\\ \hline
$k$-tripeak~\cite{wu2019towards} & $C^{ktp}_{k}$ & Triangle-based\\ \hline
At-least-$k$ clique & $C^{clique}_k$ & Clique-based\\ \hline
$k$-distance clique~\cite{alba1973graph} & $C^{dist}_k$ & Clique-based\\ \hline
$k$-VCC~\cite{torrents2015structural}&  $C^{vcc}_k$ & Connected component \\ \hline
$k$-ECC~\cite{wang2015simple}&  $C^{ecc}_k$ & Connected component\\ \hline
Alphacore~\cite{victor2021alphacore}        & $C^{alpha}_{\alpha}$ & Others\\ \hline
$k$-core-truss~\cite{li2018discovering}     & $C^{kst}_{k,\alpha}$ & Core\&Triangle-based\\ \hline
$(k,s)$-core~\cite{zhang2018discovering}       & $C^{ksc}_{k,s}$ & Core\&Triangle-based \\ \hline
SCAN~\cite{xu2007scan} & $C^{scan}_{k,\epsilon}$ & Others\\ \hline
\hline
\end{tabular}
\end{table}

We next present several definitions which are commonly used in several models. 

\begin{definition}
(\underline{Minimum degree constraint}) Given a graph $G=(V,E)$, and an integer $m$, $G$ satisfies minimum degree constraint if all the node $v\in V$ has at least $m$ neighbour nodes in $G$, i.e., $\delta(G)\geq m$. 
\end{definition}

\begin{definition}
(\underline{Maximality constraint}) 
Given a graph $G=(V,E)$ and a set of constraints $C$, a subgraph $H$ of $G$ satisfies the maximality constraint if there is no proper subgraph $H'$ of $G$ containing $H$ such that $H'$ satisfies all constraints in $C$.
\end{definition}

\begin{definition}
(\underline{Graph connectivity}) 
A graph $G=(V, E)$ is connected if there exists a path between any pair of nodes $u,v\in V$. 
\end{definition}

Next, we present a concept ``triangle'' which is utilised to identify cohesive subgraphs. In the following, we present related definitions. 

\begin{definition}
    (\underline{Triangle}) Given a graph $G=(V,E)$, a triangle in $G$, denoted as $\Delta_{u,v,w}$, is defined as a cycle of three nodes $\{u,v,w\}\subseteq V$ interconnceted by three edges $e_1=(u,v)$, $e_2=(v,w)$, $e_3=(u,w)$ $\in E$. 
\end{definition}

With the concept of ``triangle'', the strength of an edge is defined in terms of ``support''.

\begin{definition}
    (\underline{Support}) Given a graph $G=(V,E)$, a support of an edge $e$ is the number of triangles in $G$ that contains edge $e$.
\end{definition}

In the following sections, we discuss several cohesive subgraph models. We categorise them into five themes: (1) core-based models; (2) triangle-based models; (3) clique-based models; (4) connected component-based models; and (5) others. For each category, we present several cohesive subgraph models with examples.

\section{Cohesive Subgraph Models}\label{sec:models}

In this section, we present various cohesive subgraph models. Table~\ref{tab:alg_notation} provides an overview of these models, including their abbreviations and categories. We have organised the cohesive subgraph models into five distinct categories, reflecting their foundational principles.

\subsection{Core-based approaches}
In this section, we review $5$ core-based approaches~\cite{seidman1983network,zhang2018discovering,zhang2020exploring,bonchi2019distance,govindan2017k} which consider the minimum degree as a key factor to find or define a cohesive subgraph.


\begin{figure}[h]
\centering
\includegraphics[width=0.55\linewidth]{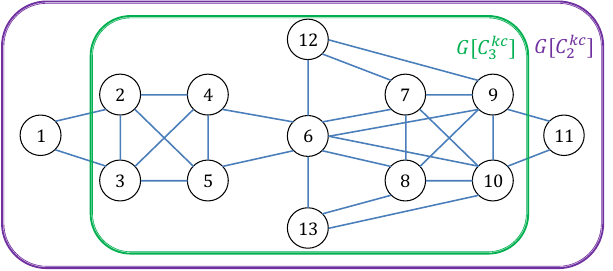}
\vspace{-0.2cm}
        \caption{Simple example for the $k$-core}
        \label{fig:kcore_example} 
\vspace{-0.2cm}
\end{figure}
\smallskip
\smallskip\noindent\underline{\spara{$\boldsymbol{k}$-core.}}
$k$-core is one of the most widely used cohesive subgraph models in networks due to its simplicity and intuitive structure. To the end users, the interpretability of the $k$-core is considered very high. The $k$-core is defined as follows. 

\begin{definition}
(\underline{\texorpdfstring{$k$-}-core}~\cite{seidman1983network}) Given a graph $G=(V,E)$ and an integer $k$, a $k$-core, denoted as $C^{kc}_k$, is a set of nodes such that
\begin{itemize}[leftmargin=*]
    \item $G[C^{kc}_k]$ satisfies maximality constraint
    \item $G[C^{kc}_k]$ satisfies minimum degree constraint with $k$
\end{itemize}
\end{definition}

Due to its straightforward and intuitive structure, the $k$-core is used for finding an initial cohesive subgraph, and serves as a foundational solution for various different problems~\cite{sozio2010community}. 
Batagelj et al.~\cite{batagelj2003m} introduced $O(|E|)$ algorithm for core decomposition. Given an integer $k$, for the $k$-core computation, we can iteratively remove a set of nodes whose current degree is less than $k$.

\begin{example}
In Figure~\ref{fig:kcore_example}, we  check the result of $k$-core $C^{kc}_2$ and $C^{kc}_3$, respectively. 
We notice that all the nodes in $G$ have at least $2$ neighbour nodes. Thus, $k$-core $C^{kc}_2$ has all the nodes in $G$. For $k=3$, we notice that nodes $1$ and $11$ do not belong to $k$-core $C^{kc}_3$ since they have only two neighbour nodes. Hence, except for the nodes $\{1,11\}$, the remaining nodes form the $k$-core $C^{kc}_3$. 
\end{example}

A direct idea to compute $k$-core is to repeatedly remove nodes whose degree is smaller than $k$ in a cascading manner. When there is no node to be removed, the remaining graph is the $k$-core. The time complexity of the $k$-core is $O(|E|)$. To check the variant of the $k$-core in different networks, please refer to these nice survey paper~\cite{malliaros2020core}.

\smallskip
\smallskip\noindent\underline{\spara{$\boldsymbol{(k,h)}$-core.}}
Bonchi et al.~\cite{bonchi2019distance} proposed a distance-based generalisation of the notion of $k$-core, namely $(k,h)$-core.

\begin{definition}
(\underline{\texorpdfstring{$(k,h)$-}-core}~\cite{bonchi2019distance}) Given a graph $G=(V,E)$, a integer $k$ and a distance threshold $h$, a $(k,h)$-core, denoted as $G[C^{khc}_{k,h}]$, is a subgraph of $G$ such that
\begin{itemize}[leftmargin=*]
    \item $G[C^{khc}_{k,h}]$ satisfies maximality constraint
    \item $G[C^{khc}_{k,h}]$ satisfies minimum degree constraint of $k$ within a distance $h$
\end{itemize}

\end{definition}

\begin{figure}[h]
\centering
\includegraphics[width=0.57\linewidth]{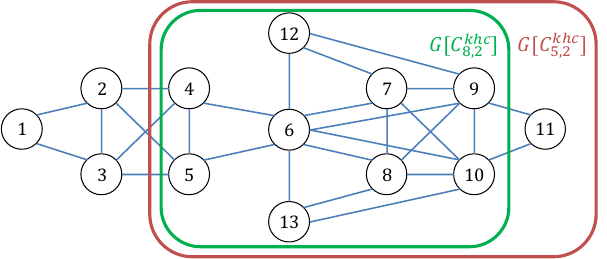}
\vspace{-0.2cm}
        \caption{Resultant subgraphs of $(k,h)$-core $C_{5,2}^{khc}$ and $C_{8,2}^{khc}$}
        \label{fig:kh_example} 
\vspace{-0.2cm}
\end{figure}

We use the example in Figure~\ref{fig:kh_example} to explain the idea of $(k,h)$-core. The subgraph in the red-coloured box represents a $(k,h)$-core $C^{khc}_{5,2}$. Similarly, the green-coloured box depicts a $(k,h)$-core $C^{khc}_{8,2}$, as each node has at least $8$ other nodes at a distance no greater than $2$ within the graph. Note that node $11$ does not belong to $(k,h)$-core $C^{khc}_{8,2}$, since there are only $7$ other nodes at a distance no greater than $2$.

In order to identify the $(k,h)$-cores, we can recursively remove the set of nodes that have fewer than $k$ other nodes at a distance no greater than $h$, until there are no nodes to be removed. The time complexity of this algorithm is $O(|V|D(D+\tilde{E}))$ where $D$ and $\tilde{E}$ are the maximum size of the subgraph induced by an $h$-neighbourhood of a node, in terms of the number of nodes and edges, respectively. 

\smallskip\noindent\underline{\spara{$\boldsymbol{(k,p)}$-core.}}
Zhang et al. \cite{zhang2020exploring} assume that each user in a cohesive subgraph is likely to be more engaged if a certain fraction of neighbours are inside the cohesive subgraph. The definition of the $(k,p)$-core is as follows.

\begin{definition}
(\underline{Fraction constraint}) Given a graph $G=(V,E)$, and a fraction $p$, a subgraph $H$ of $G$ satisfies the fraction constraint if a degree of node $v$ in $H$ divided by the degree of $v$ in $G$ is at least $p$, i.e, deg($v, H$)/deg($v, G$) $\ge$ $p$.  
\end{definition}

\begin{definition}
(\underline{\texorpdfstring{$(k,p)$-}-core}~~\cite{zhang2020exploring}) Given a graph $G=(V,E)$ an integer $k$ and a fraction threshold $p$, a $(k,p)$-core, denoted as $G[C^{kpc}_{k,p}]$, is a subgraph of $G$ such that
\begin{itemize}[leftmargin=*]
    \item $G[C^{kpc}_{k,p}]$ satisfies maximality constraint
    \item $G[C^{kpc}_{k,p}]$ satisfies minimum degree constraint with $k$
    \item $G[C^{kpc}_{k,p}]$ satisfies fraction constraint of $p$
\end{itemize}
\end{definition}

\begin{figure}[ht]
\centering
\includegraphics[width=0.55\linewidth]{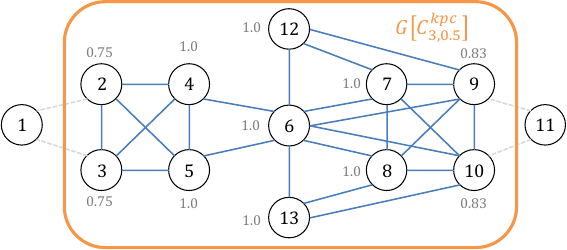}
\vspace{-0.2cm}
        \caption{Simple example for the $(k,p)$-core}
        \label{fig:kpcore_example} 
\vspace{-0.2cm}
\end{figure}

\begin{example}
In Figure~\ref{fig:kpcore_example}, $(k,p)$-core $C^{kpc}_{3,0.5}$ is presented. The fraction $p$ of each node is also marked.  It excludes nodes $1$ and $11$ since both nodes do not satisfy the minimum degree constraint. When $p > 0.75$, it returns nothing since the deletion of the nodes is applied cascadingly until there are no nodes in the remaining graph. 
\end{example}

\smallskip\noindent\underline{\textbf{$\boldsymbol{k}$-peak.}}
Recently, Govindan et al.~\cite{govindan2017k} proposed a $k$-peak decomposition problem by presenting the weakness of the $k$-core. They claimed that $k$-core might not retrieve a proper global representation of the network instead of focusing on the densest of these regions. Thus, for a better understanding of the global structure of distinct regions of a network, the authors presented a $k$-peak decomposition. To introduce the problem, $k$-shell and  $k$-contour are presented. 

\begin{definition}
(\underline{$k$-shell}) Given a graph $G=(V,E)$ and an integer $k$, $k$-shell of a graph $G$ is a set of nodes belonging to the $k$-core of $G$ but not to the $(k+1)$–core. 
\end{definition}

\begin{figure}[ht]
\centering
\includegraphics[width=0.78\linewidth]{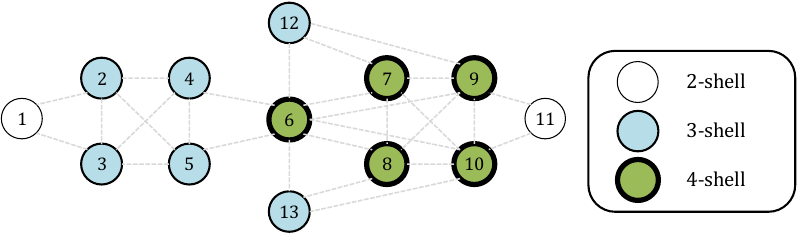}
\vspace{-0.2cm}
        \caption{$k$-shell structure}
        \label{fig:kshell} 
\vspace{-0.4cm}
\end{figure}

\begin{definition}\label{def:kcontour}
(\underline{$k$-contour}) Given a graph $G=(V,E)$ and an integer $k$, a $k$-contour, denoted as $G[T_k]$, is an induced subgraph over the maximal set $T_k$ of
nodes of $V$ such that (1) $G[T_k]$ does not contain nodes
from a higher contour, and (2) $\forall v \in T_k$, node $v$ has at least $k$ edges to other nodes
in the $G[T_k]$.
\end{definition}

The concept of $k$-contour is very similar to $k$-shell. However, it only counts the connections within the same contour.

\begin{definition}
(\underline{\texorpdfstring{$k$-}-peak decomposition}~\cite{govindan2017k}) Given a graph $G=(V,E)$ and an integer $k$, the $k$-peak, denoted as $C^{kp}_{k}$, the $k$-peak decomposition is found the assignment of each node to exactly one $k$-contour. 
\end{definition}

\begin{definition}
(\underline{\texorpdfstring{$k$-}-peak}~\cite{govindan2017k}) Given a graph $G=(V,E)$ and an integer $k$, a $k$-peak, denoted as $G[C^{kp}_k]$, is a subgraph of $G$ such that
\begin{itemize}[leftmargin=*]
    \item $G[C^{kp}_k]$ satisfies maximality constraint
    \item $G[C^{kp}_k]$ is a union of $k'$-contour, $\forall k'\geq k$
\end{itemize}
\end{definition}

\begin{example}
In Figure~\ref{fig:kshell}, we notice that the nodes $\{2,3,4,5,12,13\}$ belong to $3$-shell, but only $\{2,3,4,5\}$ belongs to $3$-contour since the nodes $12$ and $13$ connected to the nodes in $4$-shell. 
\end{example}

\begin{figure}[ht]
\centering
\includegraphics[width=0.75\linewidth]{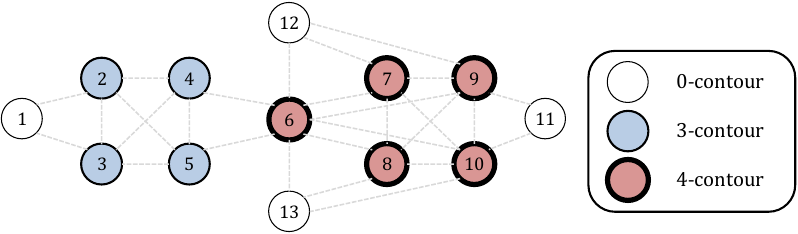}
\vspace{-0.2cm}
        \caption{Simple example for the $k$-contour}
        \label{fig:kpeak_example} 
\vspace{-0.4cm}
\end{figure}

\begin{example}
In Figure~\ref{fig:kpeak_example}, we can check which contour each node belongs to. when $k=3$, $k$-peak $C^{kp}_{3}$ returns $\{2,3,4,5,6,7,8,9,10\}$ as a result. Since the nodes $12$ and $13$ belong to $0$-contour, they cannot be in the resultant subgraph.
\end{example}

\spara{Summary and analysis.} 
In this section, we discuss the core-based model, a node-centric approach to identifying cohesive subgraphs through analysis of neighbouring structures. The $(k,h)$-core model represents a generalised version that expands the concept of neighbourhood by incorporating distance metrics. In simpler terms, a $k$-core can be considered a $(k,h)$-core where $h=1$. The fractional $k$-core, or $(k,p)$-core, introduces a fractional aspect to the neighbour structure, with the $k$-core being a special case of $(k,p)$-core where $p=0$. The $k$-peak model focuses on the connectivity among nodes that maintain consistent values post-$k$-core decomposition, leading to subgraphs characterised by heightened internal cohesion. Therefore, $k$-peak subgraphs can be viewed as subsets of $k$-core subgraphs.

\subsection{Triangle-based approaches}
In this section, we discuss several triangle-based cohesive subgraph models. 

In data mining and social network analysis fields, both counting or listing triangles in a network are fundamental problems and have numerous applications~\cite{chu2011triangle} due to the cohesive structure.

\smallskip\noindent\underline{\textbf{$\boldsymbol{k}$-truss.}} $k$-truss~\cite{cohen2008trusses} is one of the most widely used cohesive subgraph model by utilising the triangle structure to define a cohesive subgraph.

\begin{definition}
(\underline{\texorpdfstring{$k$-}-truss}~\cite{cohen2008trusses}) Given a graph $G=(V,E)$ and an integer $k$, a $k$-truss, denoted as $G[C^{kt}_k]$, is a subgraph of $G$ such that
\begin{itemize}[leftmargin=*]
    \item $G[C^{kt}_k]$ satisfies maximality constraint
    \item $\forall$ edge $e \in G[C^{kt}_k]$ has support at least $k-2$
\end{itemize}
\end{definition}

\begin{figure}[h]
\centering
\includegraphics[width=0.58\linewidth]{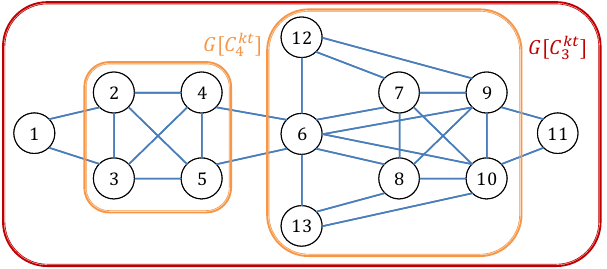}
\vspace{-0.2cm}
        \caption{Simple example for the $k$-truss}
        \label{fig:ktruss_example} 
\vspace{-0.2cm}
\end{figure}

\begin{example}
In Figure~\ref{fig:ktruss_example}, with $k=3$, the entire graph is returned by the $k$-truss returns the whole graph since every edge is involved in at least one triangle. However, when $k=4$, it returns two connected components $\{2,3,4,5\}$ and $\{6,7,8,9,10,12,13\}$ as a result. Note that there are two edges $\{4,6\}$ and $\{5,6\}$ to connect two connected components, but they involve a single triangle. Therefore, $4$-truss returns two connected components as a result. 
\end{example}

\smallskip\noindent\underline{\textbf{$\boldsymbol{k}$-tripeak.}}
Wu et al.~\cite{wu2019towards} proposed $k$-tripeak  by combining the $k$-peak~\cite{govindan2017k} and $k$-truss~\cite{cohen2008trusses}. Authors claimed that $k$-peak only accounts for the edge connection and there is an effectiveness issue due to the loose structure. Thus, they incorporated the triangle-based $k$-peak approach since the triangles in a network are considered a set of building blocks. 

\begin{definition}
    (\underline{$k$-tricontour}) Given a graph $G=(V,E)$ and an integer $k$, subgrph $H$ is tricontour of $G$ if support of $\forall e \in E \geq k-2$ and $H$ does not include edges from a higher tricontour.
\end{definition}

\begin{definition}
(\underline{\texorpdfstring{$k$-}-tripeak}~\cite{wu2019towards}) Given a graph $G=(V,E)$ and an integer $k$, a $k$-tripeak, denoted as $G[C^{ktp}_k]$, is  a subgraph of $G$ such that 
\begin{itemize}[leftmargin=*]
    \item $G[C^{ktp}_k]$ satisfies maximality constraint
    \item $G[C^{ktp}_k]$ is a union of $k'$-tricontour, $\forall k' \geq k$
\end{itemize}
\end{definition}

\begin{figure}[h]
\centering
\includegraphics[width=0.57\linewidth]{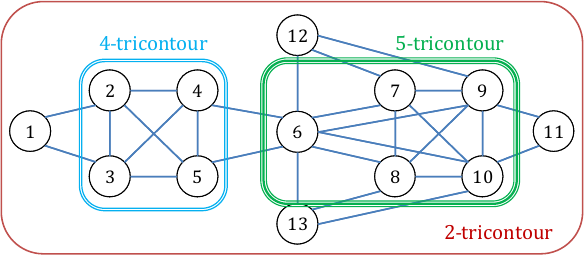}
\vspace{-0.2cm}
        \caption{Simple example for the $k$-tricontours}
        \label{fig:ktripeakcontour} 
\end{figure}

\begin{example}
In Figure~\ref{fig:main_example}, when $k=4$, $k$-tripeak $C^{ktp}_{4}$ returns two connected components : (1) $\{2,3,4,5\}$; (2) $\{6,7,8,9,10\}$. However, when $k=5$, $k$-tripeak $C^{ktp}_{5}$ returns $\{6,7,8,9,10\}$ since $\{2,3,4,5\}$ does not belongs to $k$-truss $C^{kt}_{5}$. Figure~\ref{fig:ktripeakcontour} presents $k$-tricontour of the example graph. 
\end{example}

\spara{Summary and analysis.} 
The triangle-based approach prioritises triangles as the fundamental building blocks within a network, focusing on the contribution of individual edges to the formation of triangles. The $k$-truss model is pivotal to this strategy, creating cohesive subgraphs based on the support of each edge, indicative of its involvement in triangles. The $k$-tripeak model integrates the concept of $k$-peak, applying it specifically to $k$-truss. Analogous to $k$-peak, it ensures a higher level of cohesion among edges that retain the same value following truss decomposition. Consequently, $k$-tripeak subgraphs belong to $k$-truss.

\subsection{Clique-based approaches}
In this section, we present clique-based approaches, including At-least-$k$ clique and $k$-distance clique.
A clique in a network is a subgraph where every node is directly connected to every other node in the subgraph. In other words, it is a fully connected subgraph. It has two subcategories of cliques: the $k$-size model and the $k$-distance model. A $k$-clique is a subgraph with $k$ nodes where each node is directly connected to every other node in the subgraph. A $k$-distance model is a subgraph where every pair of nodes has a distance of at most $k$ in the network. The distance between two nodes in a network is the length of the shortest path between them.

\smallskip\noindent\underline{\spara{At-least-$\boldsymbol{k}$ clique.}}
A clique \cite{luce1949method,wasserman1994social} is defined as a subgraph in which every two distinct nodes are adjacent to each other. Note that since it is NP-hard to find a clique with $k$ nodes~\cite{danisch2018listing}, we define At-least-$k$ clique. 

\begin{definition}
(\underline{At-least-\texorpdfstring{$k$-}-clique}) Given a graph $G=(V,E)$ and an integer $k$, At-least-$k$-clique, denoted as $G[C^{clique}_k]$, is a set of subgraphs of $G$ such that
\begin{itemize}[leftmargin=*]
    \item Each subgraph in $G[C^{clique}_k]$ is a clique with at least $k$ nodes
    \item Each subgraph in $G[C^{clique}_k]$ satisfies maximality constraint
\end{itemize}
\end{definition}

\begin{figure}[ht]
\centering
\includegraphics[width=0.55\linewidth]{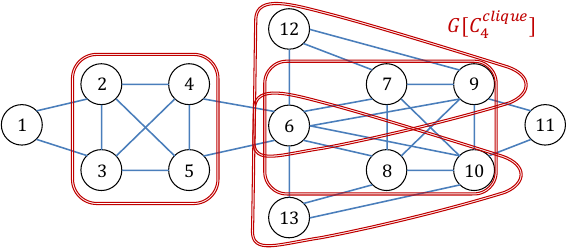}
\vspace{-0.2cm}
        \caption{At-least-$k$ clique}
        \label{fig:overksize} 
\vspace{-0.4cm}
\end{figure}

\begin{example}
In the Figure~\ref{fig:overksize}, when $k=4$, there are four At-least-$k$ cliques including 
$\{2,3,4,5\}$,  
$\{6,7,8,9,10\}$, 
$\{6,8,10,13\}$, and 
$\{6,7,9,12\}$. 
\end{example}
We use Bron-Kerbosch algorithm~\cite{bron1973algorithm} for At-least-$k$ clique. It utilises a recursive backtracking algorithm to get all maximal cliques. 
Note that the time complexity of the algorithm is $O(3^{|V|/3})$. Since we need to enumerate all the individual maximal cliques, the total time complexity is $O(|V|3^{|V|/3})$~\cite{tomita2006worst}.

\smallskip\noindent\underline{\textbf{$\boldsymbol{k}$-distance clique.}} 
We next discuss $k$-distance clique~\cite{wasserman1994social,tang2010community,alba1973graph,mokken1979cliques}. Compared with $k$ clique, it focuses on the reachability among the nodes.


\begin{definition}
(\underline{\texorpdfstring{$k$-}-distance clique}~\cite{wasserman1994social,tang2010community,alba1973graph,mokken1979cliques}) Given a graph $G=(V,E)$ and an integer $k$, $k$-distance clique, denoted as $G[C^{dist}_k]$, is a set of  subgraphs of $G$ such that
\begin{itemize}[leftmargin=*]
    \item $\forall$ pair of nodes $u, v$ within each subgraph $H$ in $G[C^{dist}_k]$, $dist(u,v,C)\leq k$  
    \item Each subgraph in $G[C^{dist}_k]$ satisfies maximality constraint
\end{itemize}
\end{definition}
Note that the distance is evaluated in the original network. Hence, it is possible that the diameter of the $k$-distance clique is larger than $k$.

\begin{property}
$dist(u,v,G) \leq dist(u,v,G[L])$ where $G[L]$ is a $k$-distance clique. 
\end{property}

\begin{figure}[ht]
\centering
\includegraphics[width=0.58\linewidth]{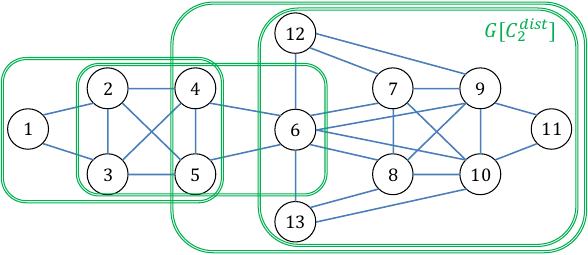}
\vspace{-0.2cm}
        \caption{Simple example for the $k$-distance clique when $k=2$}
        \label{fig:kdistance} 
\end{figure}

\begin{example}
In Figure~\ref{fig:kdistance}, we notice that when $k=2$,  there are four $k$-distance cliques including
$\{1,2,3,4,5\}$,
$\{6,7,8,9,10,11,12,13\}$,
$\{4,5,6,7,8,9,10,12,13\}$, and 
$\{2,3,4,5,6\}$. For the nodes $\{1,2,3,4,5\}$, we notice that the node $1$ is connected to all the other nodes within $2$ hops. 
\end{example}

To compute $k$-distance clique, we first generate a hyper-graph. A hyper-graph based on the distance $k$ is generated by making an edge between two nodes $i$ and $j$ if the distance between them is less than or equal to $k$. In the generated hyper-graph, we compute maximal cliques to get the result. The time complexity is the same as $k$ clique problem. 

\spara{Summary and analysis.} 
Clique-based approaches mainly aim to find a set of cliques and consider each clique as a cohesive subgraph. Thus, intuitively, its complexity is aligned with finding a set of cliques, which is known as NP-hard. The first clique-based model ``At-least-$k$ clique'' is to find a set of cliques with at least $k$ nodes. $k$-distance clique is to extend its neighbour as $k$-hops then find a set of maximal cliques. We can easily notice that At-least-$k$ clique always belongs to the $(k-1)$-core, and $h$-distance clique also belongs to $(|H|-1,h)$-core where $|H|$ is the size of $h$-distance clique.

\subsection{Connected component-based approaches}

This section introduces connected component-based cohesive subgraph models.  
The high-level idea of the connected component-based approach is to guarantee the connectivity of the remaining graph while  removing $k$ nodes (or edges) simultaneously. 
These approaches can be classified into two categories: (1) vertex-based approach named $k$-VCC, and (2) edge-based approach named $k$-ECC. We elaborate on the two types of methods in turn.

\smallskip\noindent\underline{\textbf{$\boldsymbol{k}$-VCC ($\boldsymbol{k}$-vertex connected component)}.}
$k$-VCC ($k$-vertex connected components) mainly focuses on vertex-based connectivity.

\begin{definition}
    (\underline{\texorpdfstring{$k$-}-VCC}~\cite{torrents2015structural,moody2002social})    
    Given a graph $G=(V,E)$ and integers $k$, $k$-VCC, denoted as $G[C^{vcc}_k]$, is a subgraph of $G$ such that 
\begin{itemize}[leftmargin=*]
    \item $G[C^{vcc}_k]$ satisfies maximality constraint
    \item Every connected component in $G[C^{vcc}_k]$ remains connected despite the removal of any $k-1$ nodes and disconnected upon the removal of $k$ nodes
\end{itemize}
\end{definition}

\begin{figure}[ht]
\centering
\includegraphics[width=0.55\linewidth]{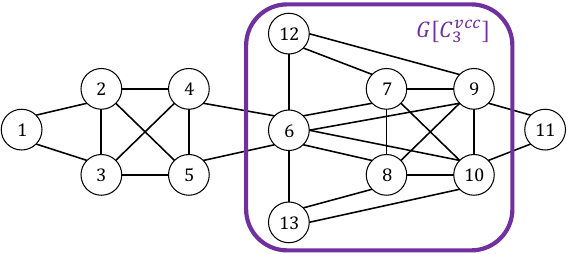}
\vspace{-0.2cm}
        \caption{Resultant subgraphs of $k$-VCC when $k=3$}
        \label{fig:kvcc2} 
\vspace{-0.4cm}
\end{figure}

\begin{example}
In Figure~\ref{fig:main_example}, when $k=3$, a connected component containing the nodes $\{6,7,8,9,10,12,13\}$ is returned. Figure~\ref{fig:kvcc2} presents the resultant subgraph. When we remove any combinations of two nodes ($(k-1)$ nodes) in Figure~\ref{fig:kvcc2}, the remaining graph is connected. 
\end{example}

\smallskip\noindent\underline{\textbf{$\boldsymbol{k}$-ECC ($\boldsymbol{k}$-edge connected component)}.}
With focusing on the edge side, $k$-ECC is proposed. 

\begin{definition}
    (\underline{\texorpdfstring{$k$-}-ECC}~\cite{wang2015simple})    
Given a graph $G=(V,E)$, and integer $k$, $k$-ECC, denoted as $G[C^{ecc}_k]$, is a subgraph of $G$ such that
\begin{itemize}[leftmargin=*]
    \item $G[C^{ecc}_k]$ satisfies maximality constraint
    \item Every connected component in $G[C^{ecc}_k]$ remains connected despite the removal of any $k-1$ edges and disconnected upon the removal of $k$ edges
\end{itemize}
\end{definition}

Note that each node cannot be disconnected to itself. Since the isolated nodes are meaningless, we return a set of cohesive subgraphs which does not consist of an isolated node.  

\begin{figure}[ht]
\centering
\includegraphics[width=0.55\linewidth]{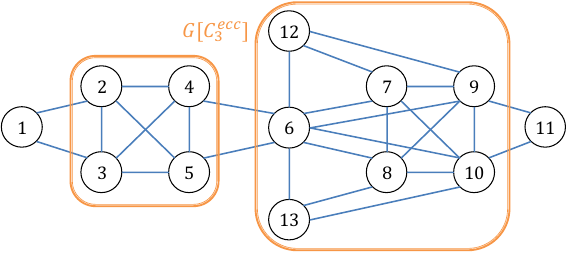}
\vspace{-0.2cm}
        \caption{Resultant subgraphs of $k$-ECC when $k=3$.  }
        \label{fig:kecc_example} 
\vspace{-0.4cm}
\end{figure}

\begin{example}
In Figure~\ref{fig:kecc_example}, it returns two connected components $\{2,3,4,5\}$ and $\{6,7,8,9,10,12,13\}$ when $k=3$. It implies that if we remove less than three edges for each subgraph, it guarantees connectivity.
\end{example}

\spara{Summary and analysis.} 
This section has discussed connected component-based approaches to cohesive subgraphs, highlighting $k$-VCC and $k$-ECC models that emphasise node and edge connectivity, respectively. $k$-VCC ensures subgraph integrity against the removal of up to $k-1$ nodes, making it ideal for applications valuing node resilience. On the other hand, $k$-ECC focuses on maintaining connectivity despite the elimination of up to $k-1$ edges, suited for scenarios prioritising uninterrupted connections. Both models offer valuable insights into network structure, providing robust tools for dissecting complex networks into cohesive components. 

\subsection{Other approaches}

\smallskip\noindent\underline{\textbf{Alphacore.}}
Victor et al.~\cite{victor2021alphacore} recently proposed a new data depth-based core decomposition technique called an Alphacore decomposition by handling multidimensional node properties, e.g., the number of edges (or triangles or cycles) that a node is a part of. It is easily applicable to directed and weighted networks by allowing multiple node properties.

In the Alphacore, \textit{data depth} is utilised to support a center-outward ordering of all observations by assigning a numerical value to each data with respect to its underlying multivariate probability distribution.

The data depth is a robust measure that quantifies the centrality of a data point within a dataset, and it is resistant to the influence of outliers. It is often used to identify the most typical or central data points in a dataset, and there are several different methods for defining it. In the Alphacore, authors utilise the Mahalanobis depth (MhD depth) as a depth function. It is one of the widely used data depth methods, and takes into account the covariance structure of the data by estimating the mean and covariance matrix of a multivariate normal distribution.

\begin{definition}
(\underline{Mahalanobis Depth(MhD Depth)}) 
Let $x\in \mathbb{R}^d$ be an observed data point. Given a $d$-variate probability distribution $F$ with mean vector $\mu_F\in \mathbb{R}^d$ and empirical covariance matrix $\sum_F\in \mathbb{R}^{d\times d}$, the MhD depth of $x$, denoted as $MhD_{\mu_F}(x)$, is defined as
\begin{align}
    MhD_{\mu_F}(x) = (1+ (x- \mu_F)^{\top} \sum_F^{-1} (x-\mu_F))^{-1}
\end{align}
where $^{\top}$ indicates matrix transpose. 
\end{definition}
The MhD depth checks the outlyingness of the point in regard to the deepest point of the distribution, and enables handling the elliptical family of distributions.
Note that $\mu_F$ can be replaced with any other point of interest.
In the Alphacore, the value is set to \textbf{0}.

\begin{definition}
(\underline{Alphacore}~\cite{victor2021alphacore}) Given a graph $G=(V,E)$ and a depth threshold $\alpha$ ,an Alphacore, denoted as $G[C^{alpha}_{\alpha}]$, is a subgraph of $G$ such that
\begin{itemize}[leftmargin=*]
    \item $G[C^{alpha}_{\alpha}]$ satisfies maximality constraint
    \item $\forall$ node $v \in G[C^{alpha}_{\alpha}]$, $v$ has at least $\alpha$ MhD Depth
\end{itemize}
\end{definition}

\begin{figure}[ht]
\centering
\includegraphics[width=0.75\linewidth]{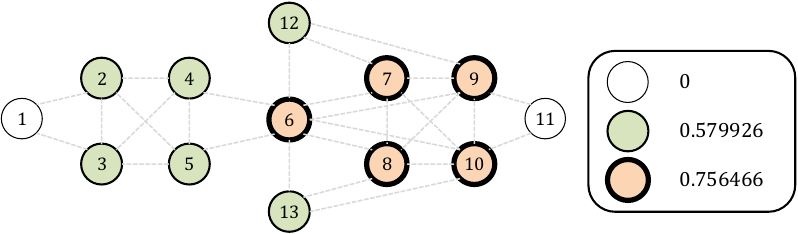}
\vspace{-0.2cm}
        \caption{Result of Alphacore}
        \label{fig:alpha} 
\vspace{-0.4cm}
\end{figure}

\begin{example}
In Figure~\ref{fig:main_example}, when we use $0 <  \alpha \leq 0.579926$, we get $11$ nodes as a result. In Figure~\ref{fig:alpha}, the union of green-coloured and pick-coloured nodes indicates the result.

Conversely, for $\alpha > 0.579926$, only the pick-coloured nodes are selected as the result. 
\end{example}

\smallskip

\noindent\underline{\textbf{$\boldsymbol{k}$-core-truss}.}
Li et al.~\cite{li2018discovering} studied $k$-core-truss problem by integrating $k$-core and $k$-truss concepts. The authors considered that when two nodes have more common neighbours, their relationship is considered as more strong. 
Furthermore, they considered relationships between nodes with high degrees to be strong connections.

\begin{definition}
(\underline{Edge degree}) Given a graph $G=(V,E)$ and an edge $e=(u,v)$ of $E$, where $u,v$ in $V$. The degree of an edge, named edge degree, is $d(e, G) = \min \{ d(u, G), d(v,G) \}$
\end{definition}

\begin{definition}
(\underline{Degree-support}) Given a graph $G=(V,E)$ and given number $\alpha$, the degree-support of an edge $e=(u,v)$ of $E$ is 
\begin{align}
    degsup_{G}(e) = \max \{sup_G(e) + 2, \alpha\cdot d(e,G) \}
\end{align}
\end{definition}

The concept of degree-support indicates the strength of the connection between two nodes. 
Due to the difference between support and degree, authors introduce a parameter $\alpha$ to balance the influence. 

\begin{definition}
(\underline{\texorpdfstring{$k$-}-core-truss}~\cite{li2018discovering}) Given a graph $G=(V,E)$, threshold $\alpha$ and $k\geq 2$, a $k$-core-truss, denoted as $G[C^{kct}_{k,\alpha}]$, is a subgraph of $G$ such that
\begin{itemize}[leftmargin=*]
    \item $G[C^{kct}_{k,\alpha}]$ satisfies maximality constraint
    \item $\forall$ edge $e \in G[C^{kct}_{k,\alpha}]$, $degsup_G(e) \geq k$ 
\end{itemize}
\end{definition}

\begin{figure}[ht]
\centering
\includegraphics[width=0.56\linewidth]{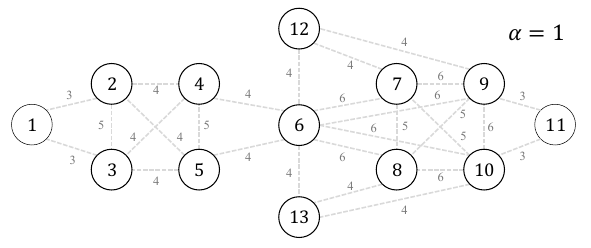}
\vspace{-0.2cm}
        \caption{Edge weightening scheme for $k$-core-truss}
        \label{fig:kcoretruss} 
\vspace{-0.4cm}
\end{figure}

\begin{example}
We use Figure~\ref{fig:main_example} to show an example. Suppose that $\alpha=1$ and $k=5$. We notice that $k$-core-truss $C^{kct}_{5, 1}$ is $\{6,7,8,9,10\}$. The nodes $1$ and $11$ do not satisfy the degree constraint since $degsup_G$ of them is $3$. Similarly, $degsup_{G\setminus \{1,11\}}$ of nodes $\{2,3,4,5,12,13\}$ is $4$. Thus, they are removed together. The $degsup$ value of the edges is presented in Figure~\ref{fig:kcoretruss}.
\end{example}

\smallskip\noindent\underline{\textbf{$\boldsymbol{(k,s)}$-core}.}
Zhang et al.~\cite{zhang2018discovering} proposed a $(k,s)$-core by unifying the $k$-truss~\cite{cohen2008trusses} and the $k$-core~\cite{seidman1983network} to identify a more cohesive subgraph while resolving the limitations of the $k$-core and $k$-truss models. Specifically, each node in the $k$-core considers its incident edges equally without considering the strength (importance), causing the $k$-core to be a promiscuous subgraph. Thus, it can be considered as ``\textit{seedbeds, within which cohesive subsets can precipitate out}''~\cite{seidman1983network}. 
In contrast, $k$-truss model is considered to have a high cohesiveness. Zhang et al. \cite{zhang2018discovering} suggested that real-world communities are node-oriented instead of edge-oriented, and it is inconsistent with reality to remove edges between the nodes in the $k$-truss only because their support is less than $k$.

To integrity the benefits of the $k$-core and $k$-truss simultaneously, Zhang et al.\cite{zhang2018discovering} first classified the edges into weak and strong ties with a user-specified threshold $s$. Specifically, an edge $e\in E$ is a \textit{strong tie} if its support $sup(e)\geq s$. Otherwise, $e$ is a \textit{weak tie}. Then, given an integer threshold $k$, a node $v$ is considered \textit{strongly engaged} if at least $k$ strong ties are incident to the node $v$, and \textit{weakly engaged} otherwise.

\begin{definition}
(\underline{\texorpdfstring{$(k,s)$-}-core}~\cite{zhang2018discovering}) Given a graph $G=(V,E)$ and integers $k,s$, $(k,s)$-core, denoted as $G[C^{ksc}_{k,s}]$, is a subgraph of $G$ such that 
\begin{itemize}[leftmargin=*]
    \item $G[C^{ksc}_{k,s}]$ satisfies maximality constraint
    \item $\forall$ node $v \in G[C^{ksc}_{k,s}]$ is strongly engaged
\end{itemize}
\end{definition}

\begin{figure}[h]
\centering
\includegraphics[width=0.8\linewidth]{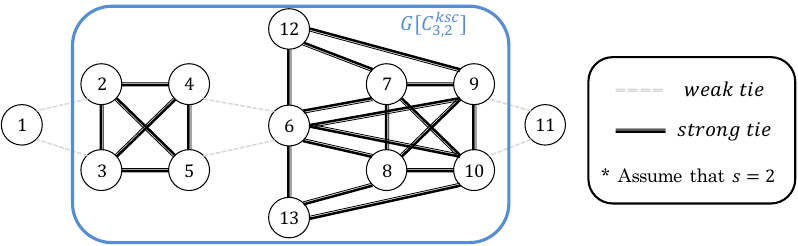}
\vspace{-0.2cm}
        \caption{$(k,s)$-core}
        \label{fig:kscore} 
\vspace{-0.4cm}
\end{figure}

\begin{example}
Reconsider the example in Figure~\ref{fig:kscore}. The $(k,s)$-core is $\{2,3,4,5,6, 7,8,9,10,12,13\}$ when $k=3$, $s=2$. Every node in this $G[C_{3,2}^{ksc}]$ has at least three edges that are involved in at least two triangles in the $G[C_{3,2}^{ksc}]$. See Figure~\ref{fig:kscore} to check the strong and weak ties when $s=2$. 
\end{example}

To identify the $(k,s)$-core, Zhang et al.~\cite{zhang2018discovering} first compute a $k'$-core in the subgraph, where $k'=max(k, s+1)$. Next, they iteratively remove a set of nodes and its incident edges if the node has insufficient strong ties until all the nodes in the remaining graph satisfy the constraint. The remaining graph is a $(k,s)$-core and is returned to end-users. The time complexity of the algorithm is $O(|E|^{1.5})$, which is the same as the time complexity of the $k$-truss computation.

\smallskip

\noindent\underline{\textbf{SCAN}.} 
SCAN~\cite{xu2007scan} is a graph-based version of the well-known DBSCAN algorithm. SCAN aims to find clusters, hubs, and outliers in a network based on the structural similarity between nodes. In SCAN, nodes are considered structurally similar if their neighbourhood structures are similar. The similarity between the two nodes is computed as follows. 

\begin{definition}
(\underline{Structural similarity}) 
Let denote $\Gamma(v)$ as $\{w \in V |(v,w)\in E\} \cup \{v\}$. Given a graph $G=(V, E)$ and two nodes $v, w\in V$, the similarity of nodes $v, w$ is defined as follows: 
\begin{align}
    \sigma(v,w) = \frac{|\Gamma(v)\cap \Gamma(w)|}{\sqrt{|\Gamma(v)||\Gamma(w)|}}    
\end{align}
\end{definition}

This SCAN algorithm operates by iteratively expanding clusters from core nodes and assigning non-core nodes to existing clusters or marking them as outliers. It distinguishes between three types of nodes: core nodes, which have sufficient structurally similar neighbours; border nodes, which have fewer than $k$ structurally similar neighbours but are reachable from core nodes and are assigned to the corresponding cluster; and outliers, which have neither enough structurally similar neighbours nor are reachable from core nodes. The formal definition of two node types (core and border) is defined as follows. 

\begin{definition}
(\underline{Core}) Given a graph $G=(V,E)$, integer $k$, and structure similarity threshold $\epsilon$, a node $v \in V$ is classified as a core node if it has at least $k$ neighbours such that $\sigma(v,w) \geq \epsilon$.
\end{definition}

\begin{definition}
(\underline{Border}) Given a graph $G=(V,E)$, integer $k$, and structural similarity threshold $\epsilon$, a node $u \in V$ is classified as a border node if it is not a core node and there exists at least one core node $v \in V$ such that $u$ is a neighbour of $v$ and the similarity $\sigma(u,v) \geq \epsilon$.  
\end{definition}

\begin{definition}
(\underline{SCAN}~\cite{xu2007scan}) Given a graph $G=(V,E)$, integer $k$ and structural similarity threshold $\epsilon$, a cluster of SCAN, denoted as $G[C^{scan}_{k,\epsilon}]$, is subgraph of $G$ such that
\begin{itemize}[leftmargin=*]
    \item $G[C^{scan}_{k,\epsilon}]$ is an induced subgraph which is formed by the union of core nodes and border nodes. 
    \item $G[C^{scan}_{k,\epsilon}]$ satisfies maximality constraint
\end{itemize}
\end{definition}

\begin{figure}[ht]
\centering
\includegraphics[width=0.55\linewidth]{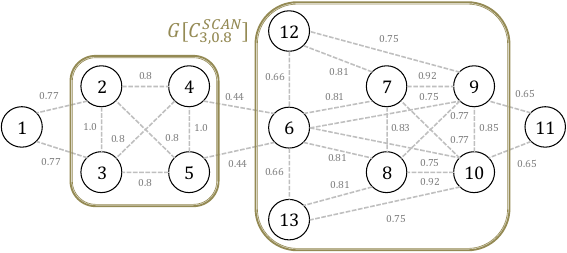}
\vspace{-0.2cm}
        \caption{Result of SCAN}
        \label{fig:scan_example} 
\vspace{-0.4cm}
\end{figure}

\begin{example}
In Figure~\ref{fig:scan_example}, when the parameters $k=3$ and $\epsilon=0.8$ are used in the SCAN algorithm, two connected components are returned: $\{2,3,4,5\}$ and $\{6,7,8,9,10,12,13\}$. The edges in the figure are labelled with the corresponding similarity scores. It can be observed that nodes 1 and 11 fail to meet the $\epsilon$ threshold, as the structural similarity of the edges $\{1,2\}$, $\{1,3\}$, $\{9,11\}$, and $\{10,11\}$ is less than $0.8$. Thus, both nodes $1$ and $11$ are considered as outliers. 
\end{example}

\spara{Summary and Analysis.}
In this section, we have discussed a set of cohesive subgraph models that do not belong to any categories that we have discussed before. Alphacore adopts a depth-based methodology to address multidimensional node attributes, leveraging the Mahalanobis depth for the identification of core structures based on depth. This approach allows for the nuanced analysis of nodes with multiple attributes, distinguishing Alphacore from traditional core detection methods. On the other hand, $k$-core-truss and $(k,s)$-core are designed to harness the strengths of both $k$-core and $k$-truss models, aiming to merge their benefits for more effective cohesive subgraph detection. SCAN, utilised in graph clustering, bears similarities to $k$-core but extends its utility by accommodating nodes' membership in multiple types, categorising them into core and border groups to define cohesive subgraphs. Note that $k$-core is equivalent to $(k,s)$-core when $s=0$. Furthermore, under the conditions $\mu=k$ and $\epsilon=0$, $k$-core can be considered a part of the results produced by the SCAN algorithm, illustrating the interconnectedness of these models in identifying cohesive structures within networks.

\subsection{Model analysis}\label{sec:categorization}

\spara{Summary of cohesive subgraph models.} Table~\ref{tab:alg-summary} summarises the cohesive subgraph models that we have studied. We note that two clique-based approaches are not in class P, while the others can be solved in polynomial time. Note that the complexity of the algorithm depends on the algorithmic process. By utilising several techniques such as precomputed indexing or pruning techniques, efficiency can be improved. When the computation methodology is not explicitly specified, we used the decomposition-based approaches.

\spara{Category-based model comparison.} In this section, we provide a comprehensive analysis of various cohesive subgraph models, categorising them based on their underlying principles and operational mechanisms.
\begin{itemize}[leftmargin=*]
\item Core-based models have an intuitive and straightforward structure, often yielding large subgraphs due to their reliance on node degree. Therefore, these approaches are better accessible for users with minimal background knowledge or serve effectively as starting points for further analysis. The major direction of the core-based models is to avoid a giant cohesive subgraph as a result. 
\item Truss-based models, in contrast, offer denser solutions but necessitate a more intricate parameter setting, potentially less straightforward for end-users. Compared to $k$-core models, users may need to adjust parameters multiple times to obtain the desired results. Due to its focus on high cohesion, truss-based models sometimes produce very small subgraphs as a result, which may not align with the needs of certain applications such as marketing target setup, recommendation, and more.
\item Clique-based models are recognised for generating highly dense, small subgraphs. Despite their intuitive structure, the strict criteria may incur prohibitive computation.
\item Connected component-based models provide subgraphs with a high degree of connectivity and robustness, even when up to $k-1$ nodes and edges are removed.  However, the selection of parameters for these models tends to be non-intuitive and overly strict, complicating the determination of suitable values.
\item Furthermore, other models like Alphacore and SCAN demand a considerable understanding from the end-user, necessitating background knowledge for their effective application.
\item Lastly, core \& triangle-based models mitigate some limitations of purely core or triangle-based approaches by furnishing stronger and denser subgraphs. Nonetheless, they encounter a similar issue to core and truss-based models regarding the difficulty of parameter setting, posing challenges in obtaining optimal results.
\end{itemize}

\begin{table}[ht]
\caption{Summary of cohesive subgraph models}
\vspace{-0.2cm}
\footnotesize
\label{tab:alg-summary}
\centering
\begin{tabular}{c|c|c|c}
\hline
\textbf{Algorithm}  & \textbf{Params} & \textbf{Time Complexity} & \textbf{Problem Class}               \\ \hline \hline
$k$-core            & $k$               & $O(|E|)$~\cite{batagelj2003m}                 & $P$                      \\ \hline

$(k,h)$-core        & $k,h$               & $O(|V|D(D+\tilde{E}))$~\cite{bonchi2019distance}                   & $P$                      \\ \hline
$(k,p)$-core        & $k,p$               & $O(|E|)$~\cite{zhang2020exploring}                   & $P$                      \\ \hline

$k$-peak            & $k$               & $O(\sqrt{|V|}(|V|+|E|))$~\cite{govindan2017k}                   & $P$                      \\ \hline

$k$-truss           & $k$               & $O(|E|^{1.5})$~\cite{cohen2008trusses}                   & $P$                      \\ \hline
$k$-tripeak           & $k$               & $O(|E|^{1.5})$~\cite{wu2019towards}                   & $P$                      \\ \hline
At-least-$k$ clique      & $k$               & $O(|V|3^{|V|/3})$~\cite{tomita2006worst}                   & $NP$ \\ \hline
$k$-distance clique & $k$               & $O(3^{|V|/3})$~\cite{tomita2006worst}                   & $NP$ \\ \hline
$k$-VCC             & $k$               & $O(|V|^4)$~\cite{torrents2015structural}                   & $P$                      \\ \hline
$k$-ECC             & $k$               & $O(|V|^4)$~\cite{torrents2015structural}                   & $P$                      \\ \hline
Alphacore           & $\alpha$               & $O(F^3 + |E||V|^2 F)$~\cite{victor2021alphacore}                   & $P$                      \\ \hline
$k$-core-truss            & $k, \alpha$               & $O(|E|^{1.5})$                   & $P$                      \\ \hline
$(k,s)$-core        & $k,s$               & $O(|E|^{1.5})$~\cite{zhang2018discovering}                   & $P$                      \\ \hline
SCAN            & $k, \varepsilon$               & $O(|E|^{1.5})$                   & $P$                      \\ \hline
\hline
\end{tabular}
\end{table}

\section{Evaluation}\label{sec:evaluation}

In this section, we evaluate various cohesive subgraph models on both synthetic and real-world networks and evaluate their performance on the community search problem. 
We classify the evaluation into two categories: local-level and global-level. The local-level evaluation treats each connected component of a cohesive subgraph as a distinct unit for evaluation while the global-level evaluation views all resultant nodes as a single result.

\subsection{Experimental setting}\label{sec:exp_setting}

\begin{table}[t]
\caption{Details of the five types of synthetic networks. Each type varies one selected parameter while anchoring the other parameters. The default parameters are $N=1,000$, $\hat{k}=20$, $maxK=200$, $\mu=0.2$, $t1=2$, $t2=1$. By default, the average clustering coefficient is automatically determined by other parameters. }
\label{tab:synthetic}

\footnotesize
\centering
\begin{tabular}{c||c|c}
\hline 
\textbf{Type} & \textbf{Selected Parameter} & \textbf{Candidate Value} $\{p1 - p4\}$\\ \hline \hline
Syn\#1 & Average degree $\hat{k}$ & $\{10, 20, 30, 40\}$ \\ \hline
Syn\#2 & Maximum degree $maxK$ & $\{100, 200, 300, 400\}$ \\ \hline
Syn\#3 & Mixing parameter $\mu$ & $\{0.2, 0.4, 0.6, 0.8\}$ \\ \hline
Syn\#4 & Minus exponent for deg. seq. $t1$ & $\{1.5, 2, 2.5, 3\}$ \\ \hline
Syn\#5 & Minus exponent for comm. size seq. $t2$ & $\{1, 1.5, 2, 2.5\}$ \\ \hline \hline
\end{tabular}
\end{table}
%
%
%
\spara{Synthetic datasets.} We use the LFR benchmark~\cite{lancichinetti2008benchmark} to generate synthetic datasets with different characteristics. The LFR benchmark takes various parameters. In this experiment, we select representative five parameters: average degree, maximum degree, mixing parameter, minus exponent for the degree sequence, and minus exponent for the community size sequence. We generate five types of synthetic graphs by varying one parameter while fixing the other four parameters. For each type of synthetic graph, we try four distinct values for the selected parameter and generate four synthetic graphs. The details of the five types of synthetic graphs are reported in Table~\ref{tab:synthetic}. For $t1$ and $t2$, we consider the value as an absolute value to avoid confusion, i.e., $1 \rightarrow 1.5$ indicates that the value is increased. 

\spara{Real-world dataset.} 
Table~\ref{tab:real-world} presents twelve real-world networks that were selected to evaluate the performance of the various cohesive subgraph models. 
We chose these networks to provide a representative sample of different types of networks from various domains, including social, e-commerce, and social media networks~\cite{yang2012defining,adamic2005political,zachary1977information,li2014efficient}.
The networks range in size from a few hundred to million nodes. For the Karate and Polblogs networks, we used the ground-truth communities provided with the datasets to evaluate the models, while for the other datasets, we evaluated the models based on their ability to identify cohesive subgraphs that correspond to meaningful structures in the network. 
Unlike the Karate and Polblogs datasets, where we used ground-truth communities for evaluation, the other datasets cannot be evaluated in the same way due to the absence of ground-truth communities. Hence, we evaluate the quality of cohesive subgraphs through various measures such as modularity, vertex density, inverse conductance, and other evaluation metrics.

\begin{table}[t]
\centering
\footnotesize
\caption{Real-world networks}
\vspace{-0.2cm}
\label{tab:real-world}
\begin{threeparttable}
    \begin{tabular}{c||c|c|c}
    \hline
    \textbf{Dataset}     & $\boldsymbol{|V|}$ & $\boldsymbol{|E|}$ & \textbf{maxC}* \\ \hline \hline
    \textbf{Karate}      & 34   & 78   & 4    \\ \hline
    \textbf{Polblogs}    & 1,224   & 16,718   & 36   \\ \hline \hline
    \textbf{YELP}      & 22,917   & 265,533   & 69    \\ \hline
    \textbf{Brightkite}      & 58,228   & 214,078   & 52    \\ \hline
    \textbf{Gowalla}      & 196,591   & 950,327   & 51    \\ \hline
    \textbf{DBLP} & 317,080   & 1,049,866   & 113    \\ \hline
    \textbf{Amazon}      & 334,863   & 925,872   & 6    \\ \hline
    \textbf{LA}      & 500,597   & 1,462,501   & 120    \\ \hline
    \textbf{NYC}      &715,605    & 2,552,315   & 157    \\ \hline
    \textbf{Weibo} & 1,019,055   & 8,245,456   & 11    \\ \hline
    \textbf{YouTube}      & 1,134,890   & 2,987,624   & 51    \\ \hline
    \textbf{LiveJournal} & 3,997,962   & 34,681,189   & 360    \\ \hline
    \hline
    \end{tabular}
    \begin{tablenotes}
    \small
    \item  *\textbf{maxC} is the maximum coreness of a network. 
    \end{tablenotes}
\end{threeparttable}
\end{table}

\spara{Evaluation metrics.} 
To evaluate the quality of cohesive subgraphs, we employ various metrics. Our evaluation of cohesive subgraphs is twofold: First, we view the cohesive subgraphs as a set of connected components, calculating measures for each and then averaging these scores. This method is denoted as \textit{local evaluation}. Alternatively, in \textit{global evaluation}, we consider a group of nodes as a subgraph regardless of connectivity. For this, we adapt the evaluation techniques described in \cite{chu2020finding}.

\spara{Global-level evaluation metrics.}
In our global-level evaluation, we evaluated performance using an aggregated measure of all connected components, forming a subgraph denoted as $H$. This approach is adapted from community scoring metrics as described by Chu et al. \cite{chu2020finding}. We use global-level evaluation metrics for real-world networks, which are often large and yield numerous subgraphs in many cohesive subgraph models. By employing global-level metrics, we can avoid unexpectedly long evaluation times.

\begin{itemize}[leftmargin=*]
    \item Average Degree: Average degree measures the typical connection of nodes in the graph, defined as the sum of node degrees in the subgraph by the total number of nodes in i.e., $\frac{2\times|E[H]|}{|V[H]|}$. A higher average degree means that, on average, the nodes in the group are more connected to each other, and consequently, it indicates a community with stronger cohesion between nodes.
    \item Cut Ratio: Cut ratio is a metric utilised to evaluate the quality of a partitioned graph by measuring the ratio of edges that exist to the number of all possible edges that exist between a set of nodes $H$ and complement of $H$, i.e., $1 -\frac{\sum_{i\in H, j\in \bar{H}} 1}{|V[H]| \times |V[G \backslash H]|}$. Cut ratio close to $1$ implies that the group has few external connections, indicating an internally cohesive group. 
    
    \item Clustering Coefficient: Clustering coefficient measures the tendency of network nodes to form clusters, and is calculated as the ratio of the number of triangles to the total number of triplets with one edge absent from the triangle. It is defined as $\frac{3\times \Delta_{u,v,w}}{\frac{d_H(d_H-1)}{2}}$, where $\Delta_{u,v,w}$ represents the set of all triangles formed by any three node $u,v$ and $w$ within $H$. A higher clustering coefficient indicates that the cohesion among a neighbours of nodes is stronger, meaning that the group is more closely connected.
    \item Edge Density: Edge density is defined as the ratio of the number of existing edges to the maximum of possible edges within a network. That is, it expressed as $\frac{2\times|E[H]|}{|V[H]|(|V[H]|-1)}$. A network characterised by high edge density indicates that it has a significant number of connections, reflecting strong interconnectivity between nodes.
    \item Inverse Conductance: Conductance is used as a metric to evaluate the level of connectivity between a specific group and external groups within the network. In this paper, we use inverse conductance, defined by the formula $1-\frac{\sum_{i\in H, j\in \bar{H}} 1}{\min(d_H, d_{\bar{H}})}$, represents the complement of conductance to have the consistency of other metrics.
    \item Average Connected Component Size: 
    Unlike other global-level metrics, Average connected component calculates the average value of the node size within each connected component. Rather than aggregating the sizes of all connected components, this metric represents the average size of the nodes in each connected component, the same as at the local-level, expressed mathematically as $\sum_{c\in C} |V[c]|/|C|$.
\end{itemize}

\spara{Local-level evaluation metrics}
In local-level evaluation, we calculate the average value of cohesive subgraphs to ensure the consistency of the evaluation metric. We measure the metric for each connected component and return the average of all results. Global-level evaluations aggregate and measure all connected components, while local-level evaluations measure the average of sum of all connected components.
Thus, the penalty of the sparse connection between disconnected  component is mitigated. We use some new notations for local-level evaluation. Let $C$ represent a set of cohesive subgraphs, with $c \in C$ indicating a specific connected subgraph. The term $l_c$ refers to the count of internal edges in subgraph $c$, and $d_c$ represents the total degree of nodes in $c$, expressed as $\sum_{u\in c} d(u, G)$, where $d(u, G)$ is the degree of node $u$ in the original graph $G$. 

\begin{itemize}[leftmargin=*]
    \item Edge Density: Edge density in local-level evaluation is the average edge density of connected components, i.e., $(\sum_{c\in C}  \frac{2|E[c]|}{(|V[c]|(|V[c]|-1))})/|C|$.
    \item Vertex Density: Vertex density in local-level evaluation is the average vertex density of connected components, i.e., $(\sum_{c\in C}\frac{|E[c]|}{|V[c]|}) / |C|$.
    \item Inverse Conductance: Inverse conductance in local-level evaluation is the average inverse conductance of connected components, i.e., $ (\sum_{c\in C}{1 - \frac{ \sum_{i\in C, j\in \bar{C}} 1 }{min(d_C, d_{\bar{C}})}}) / |C|$. 
    \item Modularity: Modularity, proposed by Newman and Girvan ~\cite{newman2004finding}, is an evaluation metric that measures the quality of subgraphs. It provides a way to evaluate how well an entire graph can be divided into cohesive and distinct subgraphs. We compute the average modularity of connected components, i.e., $(\sum_{c\in C}\frac{l_c}{|E|} - \frac{d_c^2}{4|E|^2}) / |C|$.
    \item Average Connected Component Size: Average connected component size is the average size of each connected component, the same as global-level,  i.e.,  $\sum_{c\in C} |V[c]|/|C|$.
\end{itemize}

\spara{Community Search~\cite{sozio2010community, DMCS, fang2017effective,fang2016effective}.}  We evaluate the effectiveness of the cohesive subgraph models for the community search problem by determining whether the identified subgraphs are suitable for this task. A connected component is considered a community if it contains the query node. To measure the accuracy of the identified community, we employ three representative metrics: Normalised Mutual Information (NMI), Adjusted Rand Index (ARI), and F1-score. Higher values of these metrics indicate a greater accuracy. In many community search algorithms, existing cohesive subgraphs are used as an initial solution~\cite{sozio2010community}. Therefore, it is important to note that a cohesive subgraph model with a high accuracy score does not always indicate a good model for community search, since it may imply that there is little room for improvement. However, these results can provide a rough estimate of the accuracy that a community search model might have and empirically guide us on which model could be utilised more effectively. We evaluate the models using the Karate and Polblogs datasets, which have disjoint ground-truth communities, to guarantee that the evaluation is fair and unbiased.


\spara{Algorithms.} The algorithms and parameter settings that are used in the experiments are presented in Table~\ref{tab:exp_set}. 
For $k$-core-truss, we set $\alpha=1$ by following the setting in \cite{li2018discovering}. For most cases, from $q_1$ to $q_4$, the cohesiveness level of the model becomes high. However, for several algorithms such as SCAN and $(k,p)$-core, for consistency, we have two parameter configurations. 
Selecting a proper cohesiveness threshold is challenging since every dataset has different statistics. To address this issue, we established a general parameter setting for each algorithm, which is presented in Table~\ref{tab:exp_set}. While this approach allows us to explore the performance of the models across a range of threshold values, Note that in some cases, the algorithm may return null results at certain thresholds. This is a common issue in parameter tuning for cohesive subgraph discovery problem, and it is not always feasible to tune the parameters for all datasets exhaustively. We notice that some of our experimental results may not have been well-tuned or may not have been reported in cases where the algorithm returned null results. However, we believe that our approach provides valuable insights into the relative performance of the cohesive subgraph models across different datasets and parameter settings, and we have made efforts to report and discuss all the experimental results in a transparent and unbiased manner.

\begin{table}[t]
\caption{Parameter variations $(q_1, q_2, q_3, q_4)$}
\footnotesize
\vspace{-0.2cm}
\label{tab:exp_set}
\centering
\begin{tabular}{c||c}
\hline
\textbf{Algorithm}        & \textbf{Settings $(q_1, q_2, q_3, q_4)$}                                    \\ \hline \hline
$k$-core           & $k$ = 3, 5, 7, 9                                   \\ \hline
$(k,h)$-core       & $k,h$ = (3, 2), (5, 2), (7, 2), (9,2)           \\ \hline
\multirow{2}{*}{$(k,p)$-core}      & $k,p$ = (3,0.2), (3, 0.4), (3, 0.6), (3, 0.8) \\ \cline{2-2}       & $k,p$ = (5,0.2), (5, 0.4), (5, 0.6), (5, 0.8)                \\ \hline
$k$-peak           & $k$ = 3, 5, 7, 9                                   \\ \hline
$k$-truss          & $k$ = 4, 6, 8, 10                                  \\ \hline
$k$-tripeak        & $k$ = 4, 6, 8, 10                                  \\ \hline
At-least-$k$ clique & $k$ = 3, 5, 7, 9                                   \\ \hline
$k$-distance clique & $k$ = 2, 3, 4, 5                                   \\ \hline
$k$-VCC            & $k$ = 3, 5, 7, 9                                   \\ \hline
$k$-ECC            & $k$ = 3, 5, 7, 9                                   \\ \hline
Alphacore        & $\alpha$ = 0.2, 0.4, 0.6, 0.8                    \\ \hline
$k$-core-truss     & $k,\alpha$ = (4, 1),(6, 1),(8, 1),(10, 1)        \\ \hline
\multirow{2}{*}{$(k,s)$-core}       & $k,s$ = (3, 2), (3, 3), (3, 4), (3, 5)      \\ \cline{2-2}
 & $k,s$ = (5, 2), (5, 3), (5, 4), (5, 5)    
                \\  \hline
\multirow{2}{*}{SCAN}             & $k, \varepsilon$ = (3,0.2), (3, 0.4), (3, 0.6), (3, 0.8) \\ \cline{2-2}
             & $k, \varepsilon$ = (5,0.2), (5, 0.4), (5, 0.6), (5, 0.8) \\ \hline
\hline
\end{tabular}
\end{table}

For $k$-core, $k$-truss, $k$-ECC, and $k$-VCC, we utilise the functions supported by NetworkX~\cite{hagberg2008exploring}. 
For $k$-peak, we make a wrapper of the existing implementation~\footnote{\url{https://github.com/priyagovindan/kpeak}}. 
For Alphacore, we utilise the existing python implementation~\footnote{\url{https://github.com/friedhelmvictor/alphacore}}. 
Since we use an undirected and unweighted simple graph, when the covariance matrix is not invertible, we use Moore-Penrose's pseudo-inverse~\cite{strang2006linear} via Singular Value Decomposition (SVD).

\subsection{Synthetic networks}
\begin{figure}[t]
\includegraphics[width=0.9\linewidth]{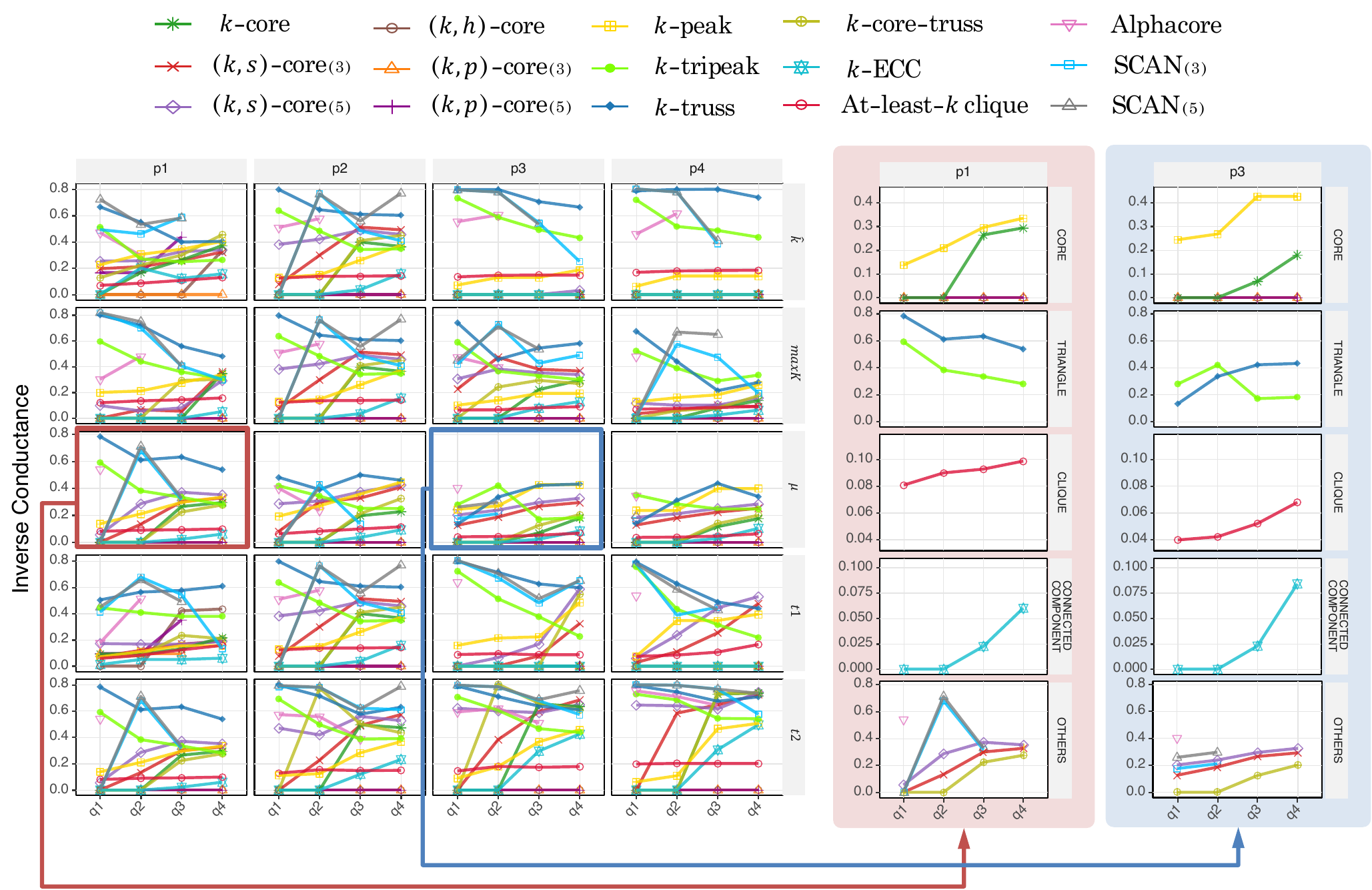}
\vspace{-0.2cm}
        \caption{Result on synthetic networks - Inv. Conductance}
        \label{fig:syn_invC_local} 
\end{figure}

\begin{figure}[t]
\includegraphics[width=0.90\linewidth]{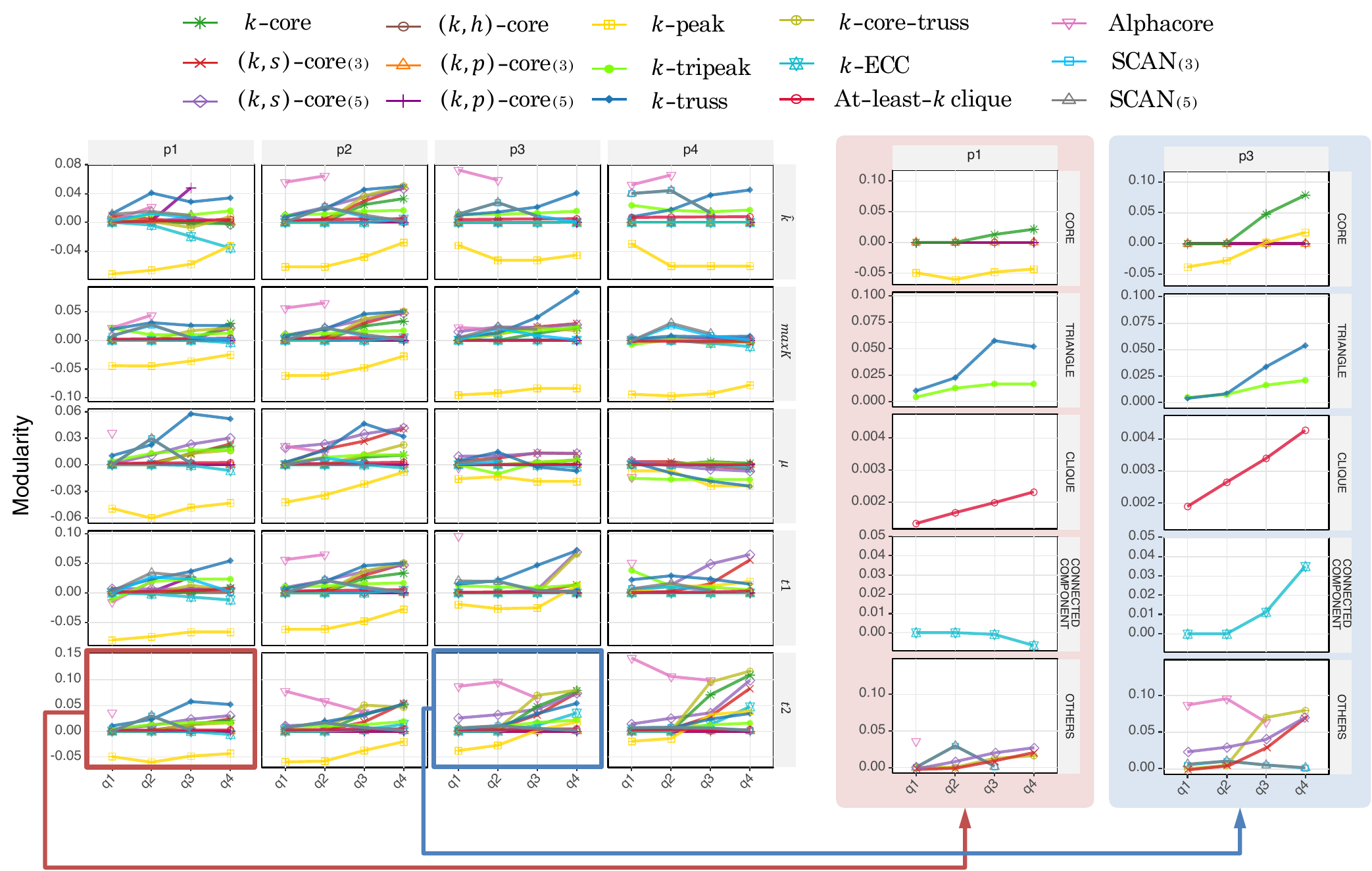}
\vspace{-0.2cm}
        \caption{Result on synthetic networks - Modularity }
        \label{fig:syn_mod_local} \vspace{-0.4cm}
\end{figure}

In this section, we evaluate the various cohesive models on the 20 synthetic network in 5 classes. The modularity, vertex density, edge density, inverse conductance, and size of the identified cohesive subgraphs are depicted in Figure~\ref{fig:syn_Eden_local} to \ref{fig:syn_size_local}, respectively. In each figure, each column of subfigures corresponds to a type of synthetic networks, while each row of subfigures corresponds to a specific value of the selected parameter in Table~\ref{tab:synthetic}. In each subfigure, the $x$-axis is the settings of cohesive subgraph models as shown in Table~\ref{tab:exp_set}. For instance, for $k$-core, $q_1$, $q_2$, $q_3$ and $q_4$ correspond to $k=3$, $k=5$, $k=7$, and $k=9$, respectively. 

\begin{figure}[t]
\includegraphics[width=0.95\linewidth]{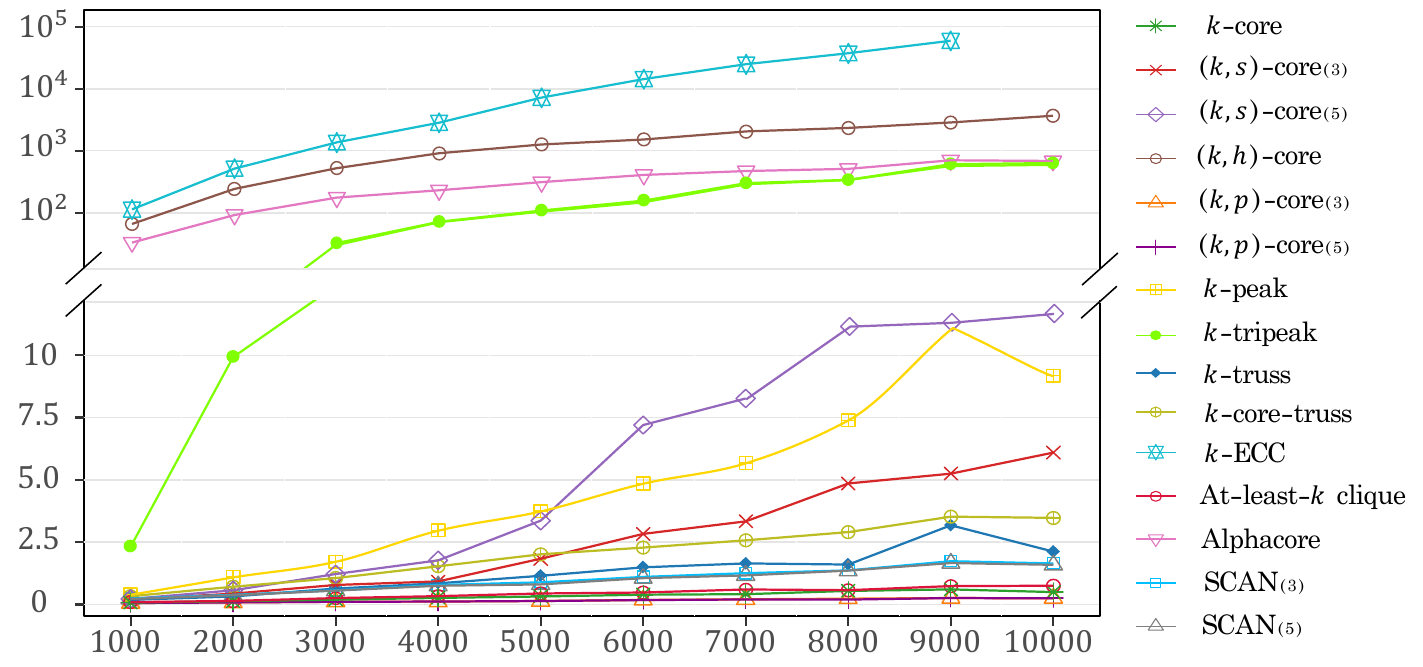}
\vspace{-0.2cm}
        \caption{Result on synthetic networks - scalability}
        \label{fig:scalability} \vspace{-0.4cm}
\end{figure}

\subsubsection{Scalability test.}
We conduct a scalability test to evaluate the performance of the various cohesive models across different node sizes in a synthetic network. The node sizes are set to range from $1,000$ to $10,000$ with increments of $1,000$ nodes each, and we measure the running time required by each cohesive model to identify the cohesive subgraphs. We select $q_3$ as a default parameter. Note that we use representative algorithms for each cohesive subgraph model. The time complexity can be checked in Table~\ref{tab:alg-summary}. 
Figure~\ref{fig:scalability} presents the results of the scalability test. The $x$-axis represents the number of nodes, and the $y$-axis represents the running time in seconds. 
Note that for some cases that required more than 24 hours, we have excluded those results. 
The $k$-VCC model is excluded due to its running time exceeding 24 hours, Similarly, $k$-distance clique is not included due to memory constraints.
For most models, the time cost increases when the node size increases. For the $k$-core, $(k,p)$-core, and SCAN models, the running time does not rise substantially with an increase in $k$. This is because of their superior time complexity compared to other cohesive subgraph models. While SCAN, $k$-truss, $(k,s)$-core, and $k$-core-truss all share the same time complexity ($O(|E|^{1.5})$, the SCAN algorithm proves to be the fastest due to its bottom-up manner.

\subsubsection{General observation.}

We present some general observations on the five types of synthetic networks.

\begin{itemize}[leftmargin=*]
    \item Syn\#1:
    As the average degree is increased, resultant cohesive subgraphs generally exhibit greater internal cohesiveness since we can expect that there are more additional internal edges in the cohesive subgraph while preserving the same ratio between internal and external edges. The increased number of internal edges may prefer to find merged cohesive subgraphs, especially, $k$-core, vertex density, and modularity. 
    \item Syn \#2: 
    When the maximum degree is increased, each node is directly connected to more nodes. This implies a graph with several giant hub nodes, potentially reducing the average distance between nodes and resulting in a more interconnected network. A higher maximum degree can reduce the average distance between nodes in the network, and this can result in a more tightly interconnected network. Thus, it makes cohesive subgraph models to identify more inaccurate results. 
    \item Syn \#3:
    A larger mixing parameter $\mu$ indicates that the tendency of clustering is decreased since the number of edges among communities is increased. Therefore, it is more difficult to identify the cohesive subgraphs when $\mu$ is larger.
    Generally, when $\mu \geq 0.5$ ($0.5$ is sometimes called critical point~\cite{george2020comparative}), it is usually considered a difficult case to find communities. Thus, when $\mu$ is extremely large, cohesive subgraph discovery approaches may fail to get a reasonable result. 
    \item Syn\#4:
    We can observe that when $t1$ increases, the degree distribution is more skewed and heavy-tailed, and the minimum degree of the nodes is increased. This higher minimum degree leads to denser node connections, complicating the differentiation between cohesive subgraphs, especially when ground-truth communities overlap.
    \item Syn\#5: We can observe that when $t2$ increases, we have a network with a higher proportion of small communities and a lower proportion of large communities. As the proportion of small communities increases, the network may become more fragmented, with more isolated groups of nodes. This could make it more challenging to understand the overall structure of the network. In general, when $t2$ decreases, finding cohesive subgraphs becomes challenging. 
\end{itemize}

\subsubsection{Experimental result analysis} 
We use several synthetic network configurations and metrics to verify the experimental results.

\begin{figure}[t]
\includegraphics[width=0.95\linewidth]{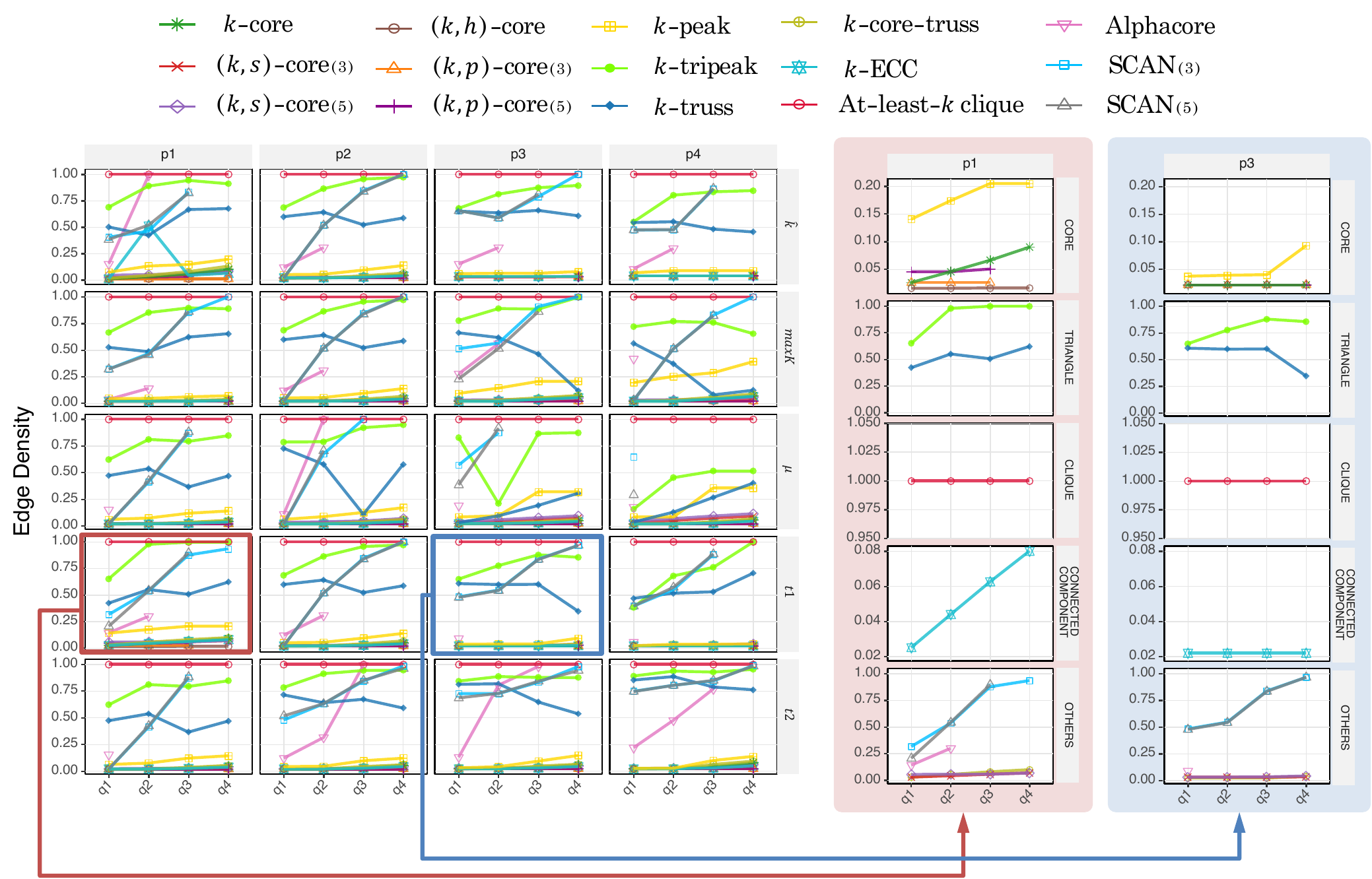}
\vspace{-0.2cm}
        \caption{Result on synthetic networks - Edge Density}
        \label{fig:syn_Eden_local} 
\end{figure}

\begin{figure}[t]
\includegraphics[width=0.9\linewidth]{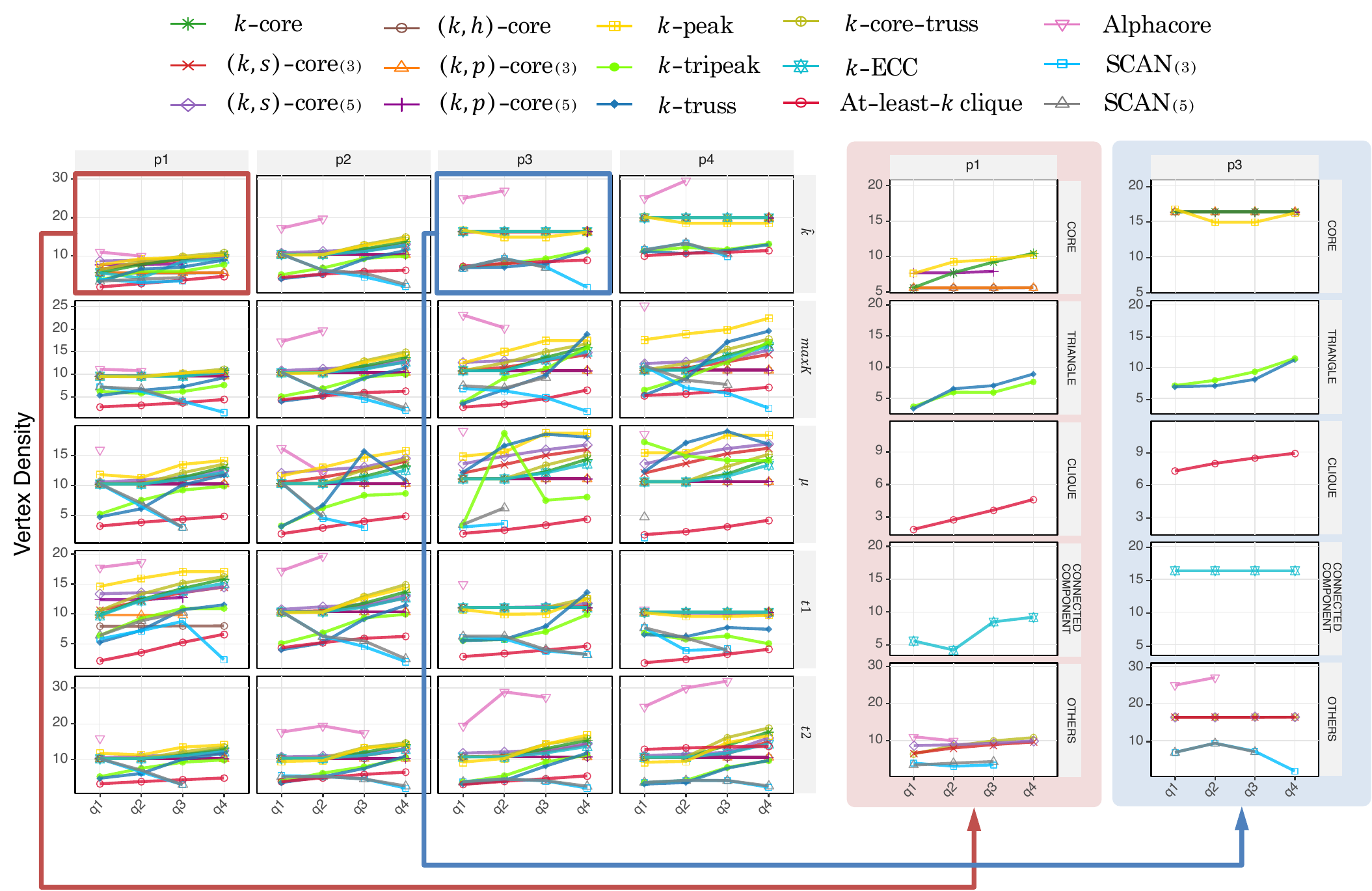}
\vspace{-0.2cm}
        \caption{Result on synthetic networks - Vertex Density}
        \label{fig:syn_Vden_local} 
\end{figure}

\spara{\textbullet{} Internal Density: Local-level Vertex Density and Edge Density.} 
In the realm of internal density—specifically focusing on local-level vertex density in Figure~\ref{fig:syn_Vden_local} and edge density in Figure~\ref{fig:syn_Eden_local}— embark on an exploration to understand how these models perform in delineating densely interconnected subgraphs. Notably, the structure of clique-based models, inherently designed to form cliques, guarantees an edge density of 1, highlighting their capacity to identify highly dense subgraphs. On the contrary, core-based models, despite their simplicity and ease of use, tend to produce larger, less dense subgraphs due to their reliance on node degrees as the primary criterion.

The analysis further reveals that models emphasising triangle formations, such as $k$-tripeak and $k$-truss, consistently exhibit higher edge densities. This observation points out the effectiveness of triangle-based strategies in capturing tightly knit cohesive structures, offering a robust mechanism to identify subgraphs characterised by dense connections. However, note that while these models excel in achieving high internal density, the $(k,s)$-core and $k$-core-truss models—a hybrid approach focusing on both node and edge criteria—sometimes show lower edge density, illustrating the trade-offs involved in balancing between different cohesive criteria.

In contrast, models like Alphacore and SCAN, although they require a deeper understanding and careful parameter tuning, have demonstrated their ability to identify cohesive subgraphs with variable densities. This adaptability, however, comes with its challenges, as maintaining consistent performance across diverse network conditions can be demanding. Due to the inherent challenge of selecting an appropriate user parameter, Alphacore, in particular, may sometimes fail to find a solution.

Through this high-level analysis, we aim to provide insights into the internal density characteristics of various cohesive subgraph models, highlighting their strengths, limitations, and suitability for different network analysis scenarios. This exploration not only aids in understanding the performance of each model but also guides future research directions in the development and application of cohesive subgraph identification methods.

\spara{\textbullet{} Internal and External Density: Local-level Inverse Conductance and Modularity.}

In evaluating the complicated structures of cohesive subgraphs, we check the internal and external densities, particularly through the lenses of local-level inverse conductance and modularity. These metrics serve as critical indicators of a subgraph's internal cohesion relative to its external separateness, offering insights into the structural cohesiveness within networks.

Inverse conductance, a measure highlighting the balance between  internal and external connections of a subgraph, reveals profound distinctions among the categorised models. Models like $(k,p)$-core and $(k,h)$-core, often return the entire graph. Thus, it returns 0 as a result. This contrasts with edge-centric models such as $k$-truss and $k$-tripeak, which stand out for their ability to retrieve highly cohesive subgraphs with minimal external connectivity, as evidenced by their elevated inverse conductance. This attribute is primarily due to their methodology, which prioritises edge relationships, and thus prefer denser, more externally isolated subgraphs.

On the modularity side, we analyse the variable performance of models under differing network cohesiveness levels. While most models show high modularity under a high cohesiveness level, delineating clear cohesive subgraph structure, they tend to converge towards lower modularity as having a lower cohesiveness level, reflecting the challenges in identifying distinct subgraphs within less cohesive environments. Notably, Alphacore consistently identifies subgraphs with notable modularity across various scenarios, although its performance may fluctuate, and hard to select proper parameters, presenting the intricacies of applying cohesive subgraph models to real-world networks.

This exploration into internal and external densities through inverse conductance and modularity not only deepens our understanding of the characteristics of different cohesive subgraph models but also presents potential directions for future research. By analysing the performance of these models in identifying the cohesive structures of a network, we set the stage for the development of more sophisticated, effective tools for network analysis, designed to meet the varied needs of diverse applications and settings.

\begin{figure}[t]
\includegraphics[width=0.90\linewidth]{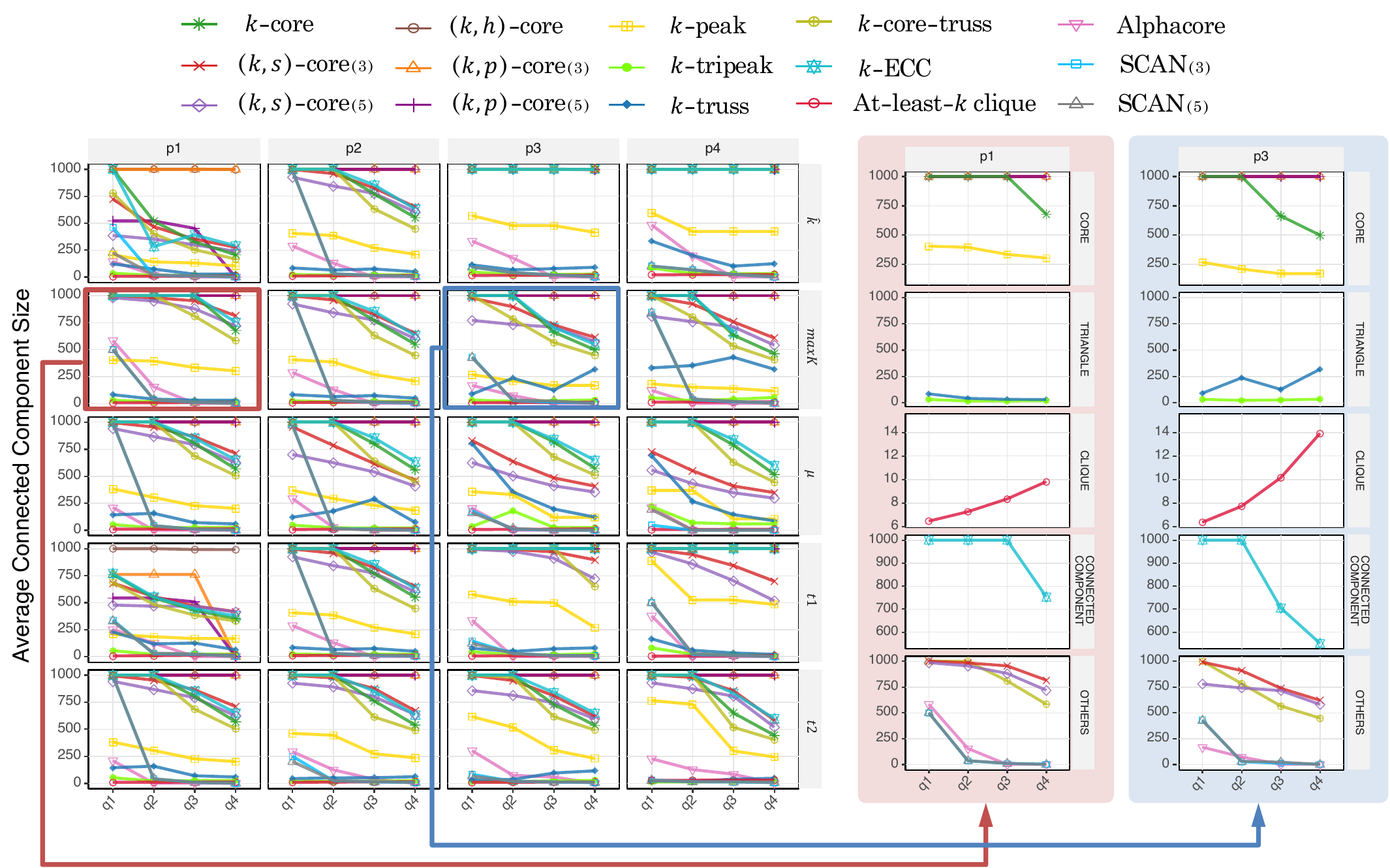}
\vspace{-0.2cm}
        \caption{Result on synthetic networks - Average Connected Component Size}
        \label{fig:syn_size_local} 
\end{figure}

\spara{\textbullet{} Local-level Average Connected Component Size.} 

By checking local-level average connected component sizes, our study highlights the performance of various cohesive subgraph models under diverse network conditions. Core-based models, such as $k$-core and its variants $(k,h)$-core and $(k,p)$-core, often yield larger subgraphs. This outcome is considered a limitation of methods based on node degrees, as they may not always capture the most cohesive or relevant community structures due to their broad inclusion criteria.

In contrast, truss-based models apply stricter cohesiveness criteria, typically identifying smaller subgraphs. This approach aims to pinpoint highly interconnected community structures, prioritising the depth of connections over the extent of the network covered. The comparison between core-based and truss-based models highlights a critical decision in the detection of cohesive subgraphs: choosing between ease of interpretation and the degree of cohesiveness.

Connected component-based methods like $k$-VCC and $k$-ECC have similar patterns with core-based approach, while clique-based models prioritise the detection of highly cohesive, albeit smaller, clusters due to their strict inclusion criteria. Innovative approaches such as Alphacore adapt multidimensional attributes for finding core structure, and models like SCAN enhance cluster identification flexibility, accommodating the complex nature of real-world networks. Bridging core and truss concepts, models like $k$-core-truss and $(k,s)$-core find a balance between interpretability and cohesion.

This investigation into how different models discern the average size of connected components not only deepens our grasp of their distinct approaches but also assists in choosing the most suitable model for particular research objectives. The decision significantly influences the analysis outcomes, whether the goal is to find easily interpretable subgraphs or to retrieve the most tightly connected subgraphs within the network.

\subsubsection{Community search}\label{sec:cs_syn}

\begin{figure}[hbt!]
\centering
\includegraphics[height=0.95\linewidth, angle=90]{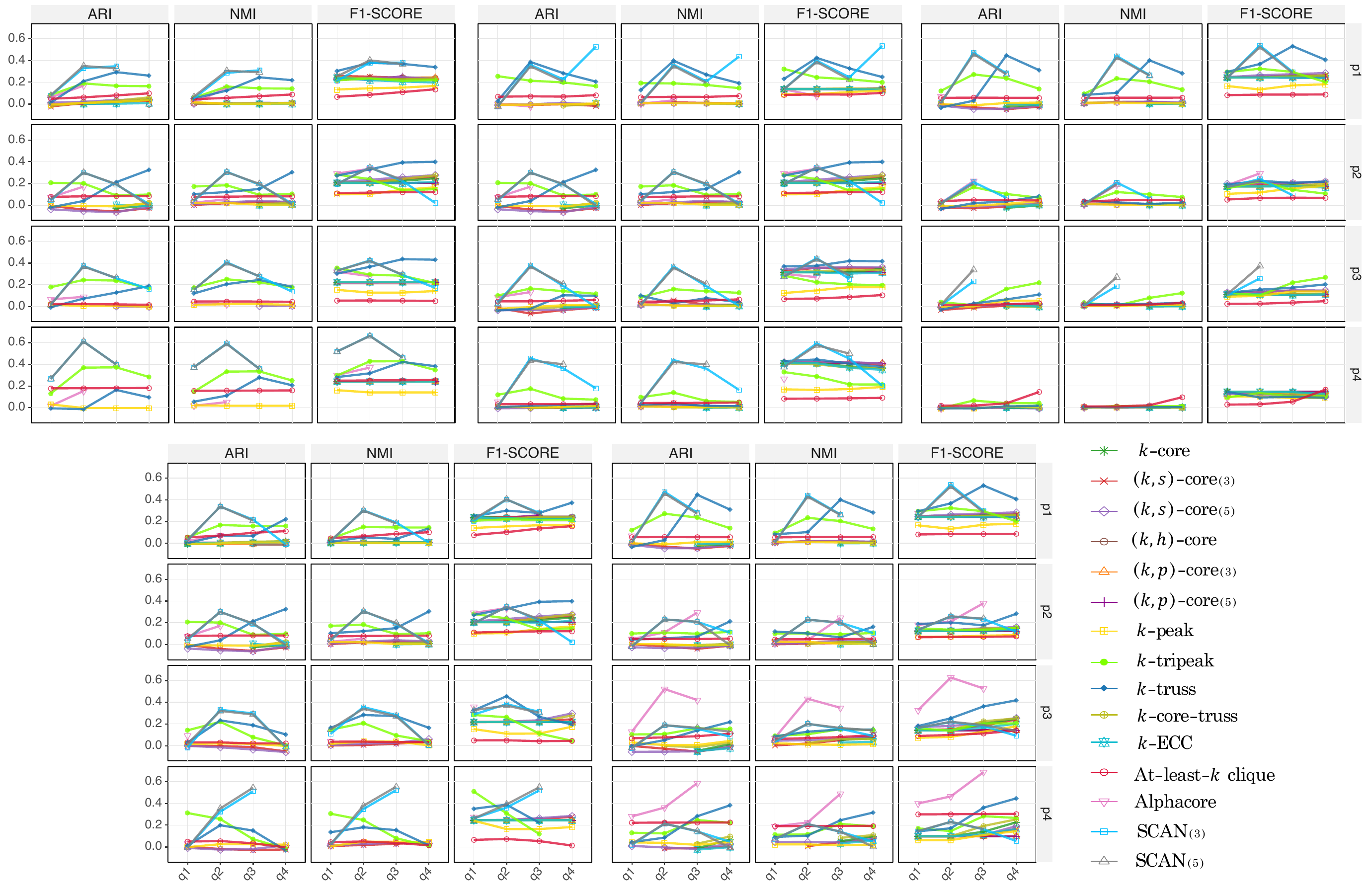}
        \caption{Accuracy of the community search problem}
\label{fig:cs_syn} 
\end{figure}

\textcolor{black}{Cohesive subgraph models can be utilised for community search problem as an initial solution. Generally, many community search models incorporate additional constraints to define communities such as diameter constraint~\cite{sozio2010community}, size constraint~\cite{sozio2010community}, keyword cohesiveness~\cite{fang2016effective}, spatial cohesiveness~\cite{fang2017effective}. Since there are numerous community search models with varying criteria, we examine the accuracy of cohesive subgraph models to determine their suitability for approaching the community search problem. 
Figure~\ref{fig:cs_syn} shows the average NMI, ARI, and F1-score values for the community search problem. After finding all cohesive subgraphs, we consider every node within an identified subgraph as a query node. We regard the connected component that contains the query node as the identified cohesive subgraph for the community search problem. Then, consider the community search problem as binary classification~\cite{DMCS} to check the effectiveness. We observe that cohesive subgraph models including Alphacore, $k$-tripeak, $k$-truss, and SCAN return more accurate results compared with other models. Since SCAN is a graph clustering algorithm, it returns a high-quality community. $k$-tripeak and $k$-truss achieve high accuracy since truss-based approaches normally return more cohesive subgraphs compared with other approaches. 
Note that returning a high accuracy score does not reflect the suitability of the cohesive subgraph models for the community search problem. This is because a high accuracy score may imply that the cohesive subgraph model filters out several nodes to satisfy the specific constraints. If we use the cohesive subgraph models as initialisation, it might have less chance for community search approaches to improve the quality.}

\begin{figure*}[t]
\centering
\begin{subfigure}{0.65\linewidth}
\centering
    \includegraphics[width=0.99\linewidth]{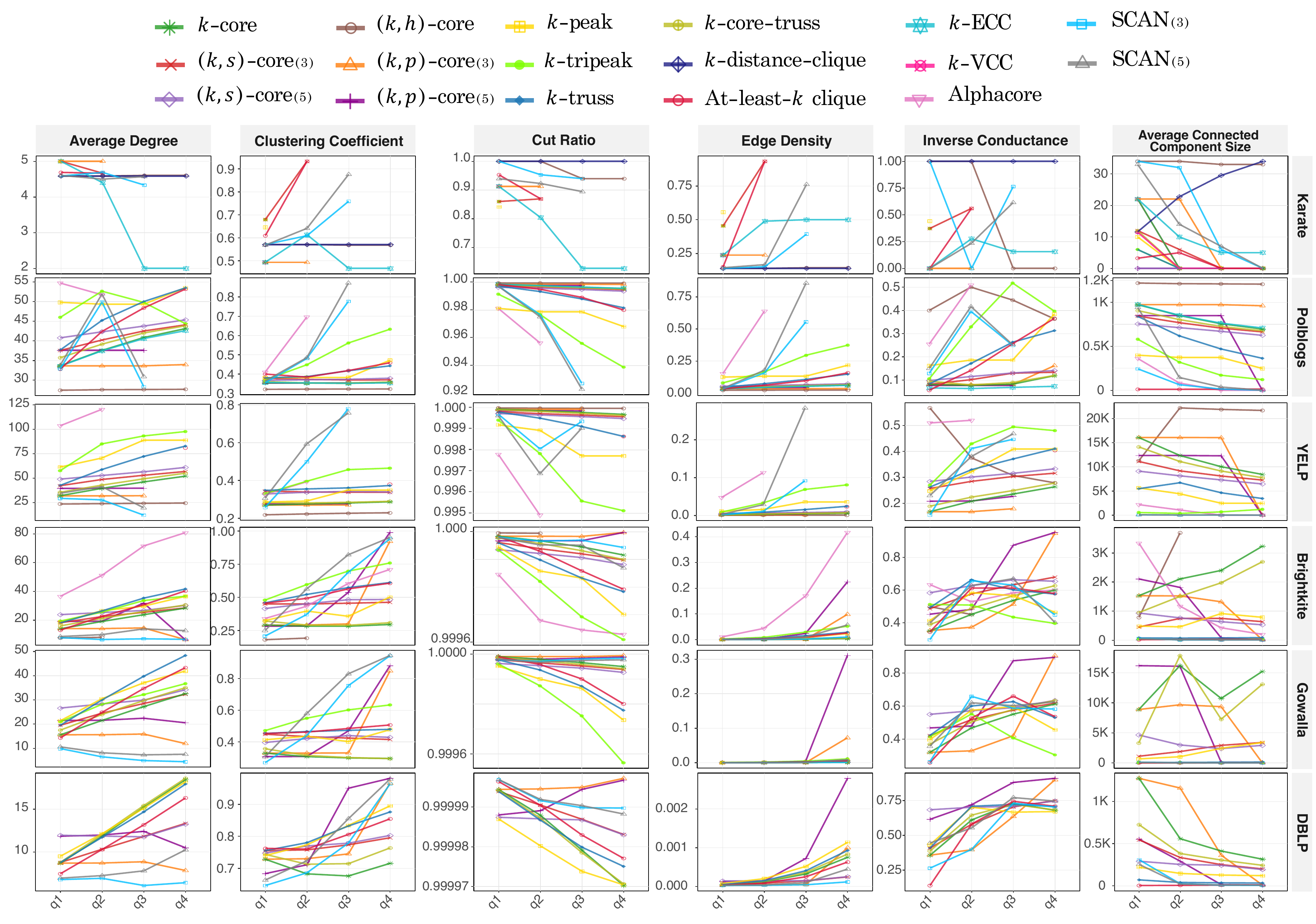}
    \label{fig:real_statistics1}
\end{subfigure}
\begin{subfigure}{0.65\linewidth}
\centering
\includegraphics[width=0.99\linewidth]{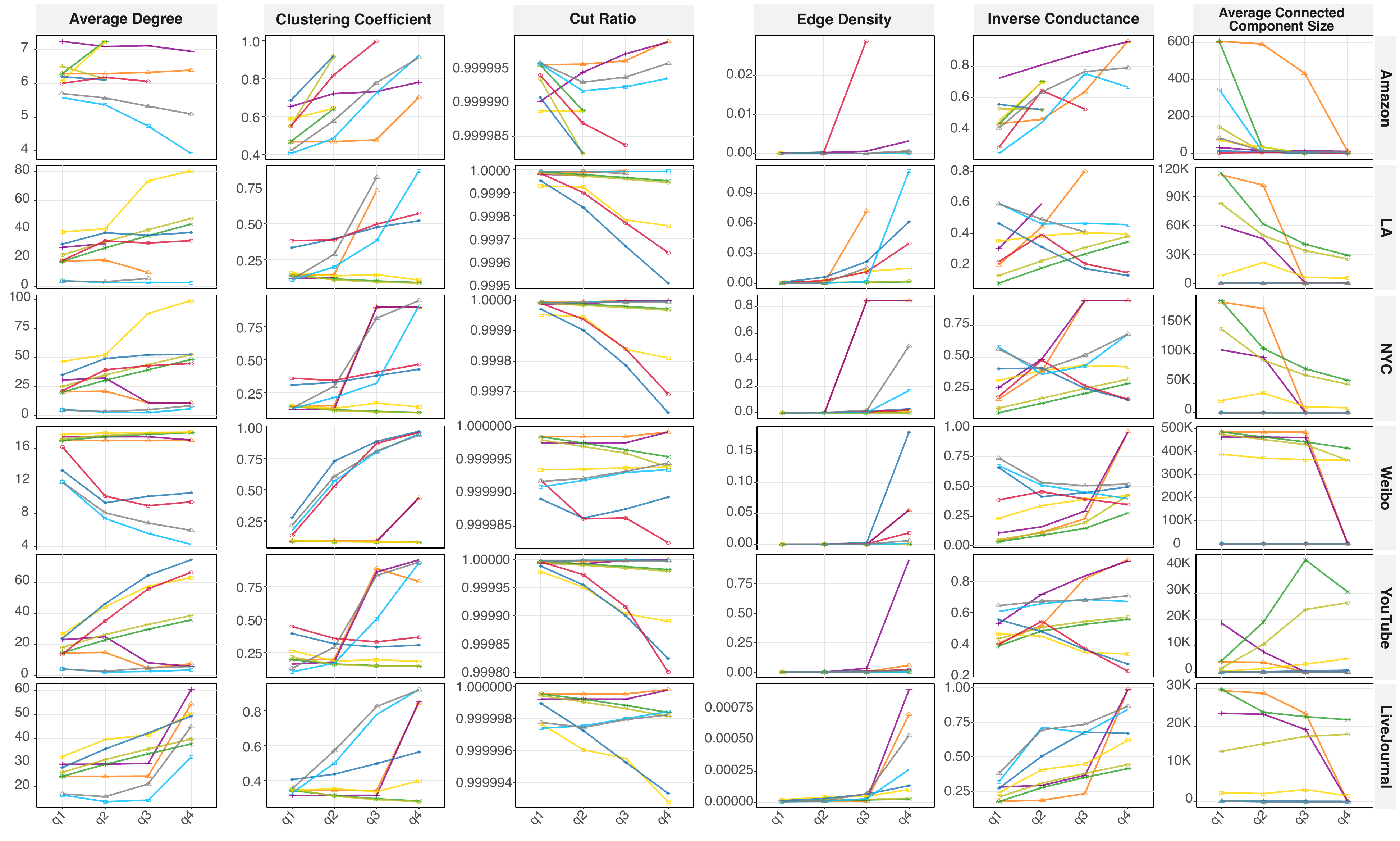}
    \label{fig:real_statistics2}
\end{subfigure}
\caption{Result on real-world networks}
\label{fig:real_statistics}
\end{figure*}

\subsection{Real-world networks}

\subsubsection{Experimental result analysis} 
We use real-world networks to verify the experimental results. Note that each network has different statistics, and thus selecting a proper parameter for each network configuration is challenging. Hence, we use the same parameter settings which can be checked in Table~\ref{tab:exp_set}. All the results are presented in Figure~\ref{fig:real_statistics}. Note that to evaluate the cohesive models, we utilise the global metrics which are presented in Table~\ref{sec:exp_setting}.

\spara{Global-level Average Degree.}
In real-world datasets, dataset size does not affect the average degree of the identified cohesive subgraphs. 
This seems to be attributed to the inherent nature of real-world data, where the complexity and diversity of connections remain consistent regardless of dataset size. 
In general, the Alphacore, $k$-truss, $k$-peak, and $(k,s)$-core models exhibit a high average degree. This might be due to their algorithms focusing on dense cohesive subgraph, which inherently leads to higher average degrees. Moreover, SCAN and $(k,p)$-core tend to have lower average degrees. 
Unlike SCAN and $k$-peak, models like $k$-core and $k$-truss consider an absolute value of the cohesiveness parameter, identifying cohesive subgraphs with a fixed $k$ value. For example, even for large graphs, setting a high $k$ value can allow these models to identify a large subgraph. However, models such as SCAN and $(k,p)$-core tend to be less scalable than models that use absolute cohesion values because they consider ratios, which are relative values. 

\spara{Global-level Clustering Coefficient.}
The clustering coefficient is used as a metric to indicate cohesiveness within a network. Compared to other models, SCAN and $(k,p)$-core exhibit a relatively high clustering coefficient. Furthermore, the triangle-based $k$-truss model, known for its strong cohesion, and the At-least-$k$ clique model, which connects to every node within a community, both demonstrate notably high clustering coefficients. In the case of SCAN and the $(k,p)$-core model, as cohesiveness increases, there is a discernible trend of the clustering coefficient steeply rising. This is attributed to the consideration of relative ratios represented by the parameters $p$ and $\epsilon$.
Conversely, the $k$-truss and At-least-$k$ clique models are known for their relatively high cohesion compared to other models. However, due to the lower values of model parameters, their clustering coefficients tend to increase more slowly when compared to models that use ratio constraints.

\spara{Global-level Cut Ratio.} 
Several models performed significantly better on cut ratio, making it difficult to determine trends. It is observed that most models have a high cut ratio: a cut ratio closer to 1 indicates a high-quality community with fewer external connections. As the scale of the dataset grows, the cut ratio for $k$-truss, At-least-$k$ clique, and $k$-peak models decreases. The cut ratio of the $k$-tripeak model decreases more significantly compared to other models as the dataset size increases. However, when dealing with extremely large datasets, it fails to identify cohesive subgraphs. At-least-$k$ clique has strict connectivity requirements due to the nature of cliques. The cut ratio may be comparatively low due to the condition that any node cannot be included in the clique even if it has high connectivity unless it is connected to all nodes. $k$-tripeak and $k$-peak models also have high requirements, so they cannot be connected even with high coreness and trussness, so they have a lower cut ratio. $k$-truss also has relatively higher requirements than the core-based model.

\spara{Global-level Edge Density.} Generally, increasing the cohesiveness level causes the model to find cohesive subgraphs with high edge density, and the cohesiveness level tends to increase as the number of internal connections within a community increases. At the local-level, the edge density is measured by averaging across each connected component. Therefore, the At-least-$k$ clique model consistently achieves an edge density of 1. However, at the global level, where all connected components are aggregated into a single resultant subgraph, this model does not invariably maintain an edge density of 1. Despite this, a comparison between the models shows that the edge density of the At-least-$k$ clique model is relatively high.  As the dataset scales larger, the cohesive subgraphs identified by $(k,p)$-core (rather than models like Alphacore and SCAN) have a relatively high edge density when the cohesiveness level is high. Note that for $(k,p)$-core, it identifies cohesive subgraphs with relatively small cut ratios and high edge density. This implies that compared to models that constrain an absolute value of $k$, $(k,p)$-core, a model that considers ratios, can have a higher edge density as the size of the network increases, as the threshold for the number of edges to satisfy $p\%$ increases. 

\spara{Global-level Inverse Conductance.} Generally, we observe that as the cohesion parameter increases, the inverse conductance also increases. As the size of the dataset expands, the inverse conductivity of the $(k,p)$-core becomes more and more pronounced. In particular, the inverse conductivity increases significantly as the parameter settings of the model increase compared to other models. This is because increasing the parameter $p$ results in more robust connections. Also, increasing the parameter $k$ returns more dense subgraphs, so larger $k$ performs better. On the other hand, $k$-core has much lower inverse conductance as the size of the dataset increases. This is because $k$-core is a simple and weakly constrainted model compared to other models, so it returns a large subgraph containing many nodes, which can be less cohesive. 

\spara{Global-level Average Connected Component Size.} In small datasets, the $(k,h)$-core consistently yields the largest cohesive subgraph due to its characteristics. Due to the scalability issue of $(k,h)$-core, it fails to identify a cohesive subgraph in large-sized networks within a reasonable time. Additionally, it is observed that as the cohesiveness parameter increases, the size of the cohesive subgraphs identified by the model decreases. In some datasets including Brightkite, Gowalla, and Youtube, models such as $k$-core exhibit larger values of size as the cohesiveness parameter $k$ increases. This means that in these datasets, the model returns a larger number of nodes for smaller values of $k$, and conversely, it returns a smaller value of size for smaller values of $k$ if there are more overlapping nodes and more subgraphs containing them. Also, even if $k$ is large, size may increase if there are fewer large cohesive subgraphs. As the dataset size increases, we observe an increase in the sizes of the $k$-core, $k$-core-truss, and $(k,p)$-core. Notably, triangle-based approaches typically produce small-sized results.

\begin{figure}[t]
\centering
\includegraphics[width=0.95\linewidth]{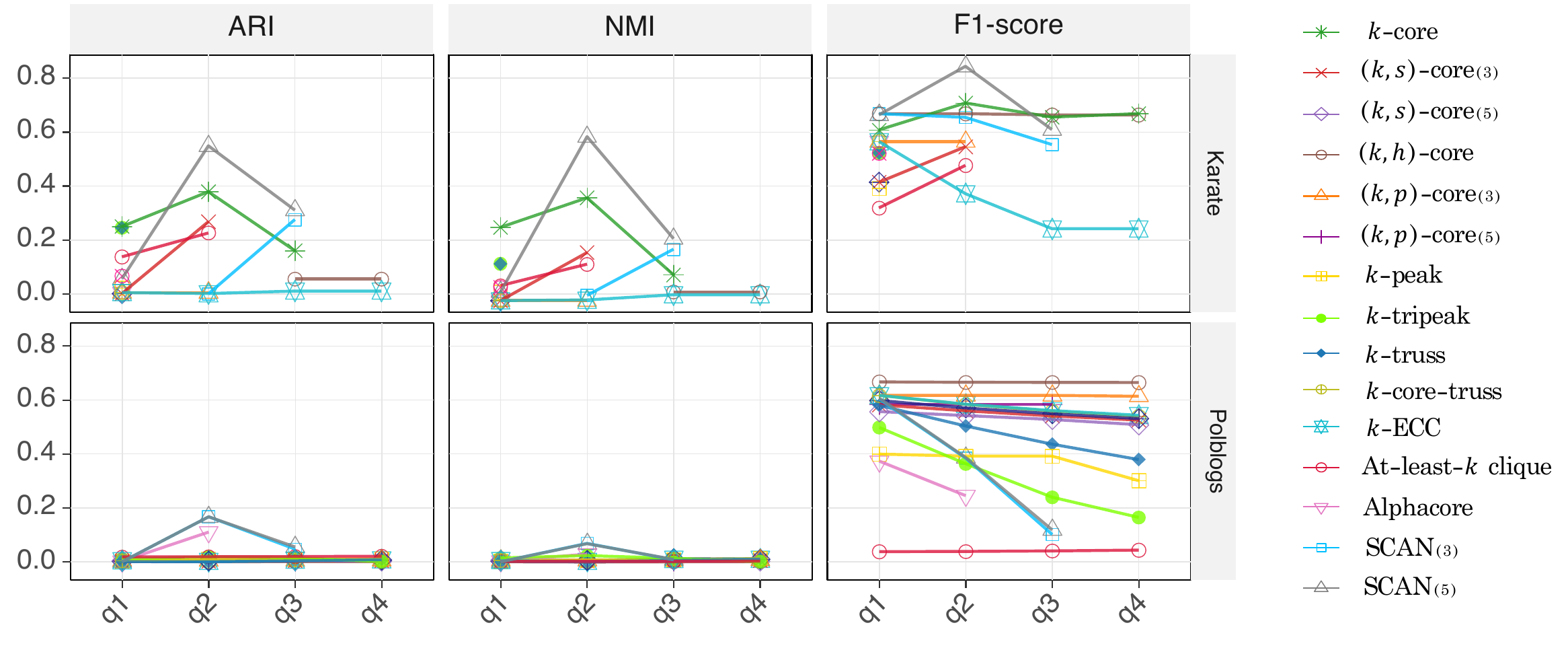}
\vspace{-0.2cm}
        \caption{Accuracy of the community search problem (Karate and Polblogs networks, Same legend with Figure~\ref{fig:real_statistics}). ARI, NMI and F1-score are the average value.   }
        \label{fig:cs_real}
\vspace{-0.4cm}
\end{figure}

\subsubsection{Community search}
In Section~\ref{sec:cs_syn}, we present a case study on the community search problem utilising the LFR benchmark dataset~\cite{lancichinetti2008benchmark}. In Figure~\ref{fig:cs_real}, we present the accuracy of the cohesive subgraph models for the community search problem on two real-world networks with ground-truth communities. We observe that the SCAN algorithm consistently outperforms other algorithms, as the goal of the SCAN algorithm is to identify community structure. We also observe that Alphacore, $k$-distance clique, $k$-core, and $(k,s)$-core return comparable results. For the SCAN, we observe that the outperformed models return relatively small-sized cohesive subgraphs as a result. This implies that returning small-sized communities (See Figure~\ref{fig:syn_size_local}) generally results in more accurate results. However, they still face challenges in selecting proper cohesiveness levels.

\subsection{Result analysis}

In this Section, we present experimental results evaluating the effectiveness and efficiency of cohesive subgraph models on both synthetic and real-world networks. In synthetic networks, we varied several user parameters and evaluated various subgraph evaluation metrics while varying the cohesiveness level of each cohesive subgraph model. Selecting an appropriate cohesiveness level is still a challenging and open problem and entails several limitations. However, through these processes, several interesting observations are identified. It shows that Alphacore, SCAN consistently produced more cohesive results compared to other cohesive subgraph models. This is because SCAN is inherently designed to consider both the internal and external edges of the graph simultaneously, excluding outliers. Our observations also show that truss-based and combined cohesive subgraph models consistently produced more cohesive results compared to core-based approaches. Core-based approaches generally typically identified larger results due to their inherent characteristics~\cite{cohen2008trusses,seidman1983network}.  
In real-world networks, we evaluated the performance of cohesive subgraph models for a downstream task. Our results indicated that truss-based approaches performed well in both tasks, producing high-quality communities. For the community search task, graph clustering-based approaches performed the best since their purpose is similar to that of the community search problem. It is worth noting that for a downstream task, we directly utilised the cohesive subgraph models. Since many models can be used as an initial solution for a specific downstream task, we suggest that to set the lower-bound, truss-based or combined approaches are preferred. However, for finding an approximated solution, utilising cohesive subgraph models which prefers to return large-sized solutions such as core-based approaches can be an option since there might be some room to improve the quality.

\section{CHALLENGES AND FUTURE RESEARCH}\label{sec:challenges}
Our study provides a comprehensive evaluation of existing cohesive subgraph models, but there are still some challenges for future research that need to be addressed. 

Evaluating cohesive subgraph models can be challenging due to the lack of standardised evaluation metrics. Although we used a combination of metrics to evaluate the proposed models, there is a need for an agreed-upon standard for evaluating cohesive subgraphs, which can help compare results across different studies. Additionally, there is a need for more realistic datasets that can capture the complexity of real-world networks to better understand the performance of cohesive subgraph models in practical settings.

While it may not be possible to definitively conclude which cohesive subgraph is universally superior or inferior, our experiments and analyses suggest customised recommendations based on specific needs. For scenarios where interpretability is a key factor, core-based methods can be recommended. In contrast, when subgraph cohesion is the primary concern, truss-based methods are more suitable. For situations requiring a balanced approach, employing a combination of core and truss-based methods is recommended. This recommendation arises from the inherent trade-off between interpretability and cohesion within the domain, highlighting the importance of prioritising one aspect over the other or finding a balanced compromise when designing new cohesive subgraph models.

A promising area for future research is exploring the use of cohesive subgraphs in other applications beyond marketing strategies and fraud detection, such as social network analysis for detecting influential spreaders, network visualisation, keyword extraction, and biology. Investigating the interpretability of cohesive subgraphs can also help to better understand the underlying structure and characteristics of complex networks. 

In conclusion, while our study provides valuable insights into the performance of existing cohesive subgraph models, there are still several challenges and areas for future research. Addressing these challenges and exploring new applications and interpretations of cohesive subgraphs can continue to advance the field of social network analysis and graph data management.

\section{CONCLUSION}\label{sec:conclusion}
This paper evaluated $14$ cohesive subgraph models through qualitative experiments on synthetic networks and real-world networks and investigated their performance in downstream tasks such as community search. This study offers meaningful observations and insights into the characteristics of these models, as well as their advantages and limitations. In addition to understanding the principles behind these cohesive subgraph models — objectives, methodologies, and the patterns discerned through various datasets — serves as a foundational basis for developing or expanding these models in diverse environments.

\section*{Acknowledge}
This work was supported by Institute of Information \& communications Technology Planning \& Evaluation(IITP) grant funded by the Korea government(MSIT)(No.2020-0-01336, Artificial Intelligence graduate school support (UNIST)), the 2022 Research Fund (1.220138.01) of UNIST, National Research Foundation of Korea (NRF) grant funded by the Korea government (MSIT) (No. RS-2023-00214065), the National Research Foundation of Korea(NRF) grant funded by the Korea government(MSIT). (No. RS-2023-00277907), and grants (No. 62394333, No. U22A2099, No. 62376028) from the National Natural Science Foundation of China.


\bibliographystyle{ACM-Reference-Format}
\bibliography{98_bib}

\clearpage


\end{document}